\documentclass[twocolumn]{aastex61}

\received{June, 2018}
\revised{July, 2018}
\accepted{July 4, 2018}
\submitjournal{ApJ}

\shorttitle{Radio observations of SN 2004dj}
\shortauthors{Nayana et al.}

\begin{document}

\title{Long term behaviour of a Type IIP supernova SN\,2004dj in the radio bands}

\correspondingauthor{Nayana A J}
\email{nayana@ncra.tifr.res.in}

\author[0000-0002-0844-6563]{Nayana A. J.}
\affil{National Centre for Radio Astrophysics, Tata Institute of Fundamental Research, PO Box 3, Pune, 411007, India}

\author{Poonam Chandra}
\affiliation{National Centre for Radio Astrophysics, Tata Institute of Fundamental Research, PO Box 3, Pune, 411007, India}
\affiliation{Department of Astronomy, Stockholm University, AlbaNova, SE-106 91 Stockholm, Sweden}

\author{Alak K. ray}
\affiliation{Homi Bhabha Centre for Science Education, 
Tata Institute of Fundamental Research, Mankhurd, Mumbai, 400088 India}

\begin{abstract}
We present radio observations and modelling of one of the nearest and brightest Type IIP supernova SN\,2004dj exploded in the galaxy NGC 2403 at a distance of $\sim$ 3.5 Mpc. Our observations span a wide frequency and temporal range of 0.24 - 43 GHz and $\sim 1$ day to 12 years since the discovery. We model the radio light curves and spectra with the synchrotron emission. We estimate the mass-loss rate of the progenitor star to be $\dot{M}$ $\sim$ 1 $\times$ 10$^{-6}$ M$_{\odot}\, \rm yr^{-1}$ for a wind speed of 10 km\,s$^{-1}$. We calculate the radio spectral indices using 1.06, 1.40, 5.00 and 8.46 GHz flux density measurements at multiple epochs. We witness steepening in the spectral index values for an extended period predominantly at higher frequencies. We explain this as a signature of electron cooling happening at the supernova shock in the plateau phase of the supernova. We estimate the cooling timescales for inverse Compton cooling and synchrotron cooling and find that inverse Compton cooling is the dominant cooling process.
\end{abstract}


\section{Introduction} \label{sec:intro}

Massive stars (M $>$ 8M$_{\odot}$) end their lives in spectacular explosions called core-collapse supernovae (SNe). Type II SNe are explosions of massive stars that retain their hydrogen envelope at the time of explosions and show copious hydrogen emission lines in their optical spectra \citep{filippenko1997}. Type IIP SNe is a sub-class of Type II SNe characterized by a pronounced plateau of nearly constant luminosity in the optical light curve following a maximum that lasts for $80-120$ days post explosion \citep{nadyozhin2003}. The plateau in their optical light curves is attributed to an extended hydrogen envelope intact to the star during explosion \citep{grasberg1971,falk1977,smartt2009b}. In a volume limited sample ($<$ 60 Mpc) of all core-collapse SNe, 48\% is comprised by SNe IIP \citep{smith2011}. Thus, SNe IIP is the most commonly observed variety of core-collapse SNe in the local universe and hence likely to be the most common evolutionary path in the end stages of the life of massive stars.

Several lines of evidence from stellar evolution models and supernova (SN) light curve models suggest that the progenitors of SNe IIP are red-supergiants \citep[RSGs;][]{chevalier2006b}. The pre-supernova radius estimated from the plateau brightness and duration also falls in the typical RSG stellar radius \citep[$10^{2-3} R_{\odot}$;][]{grasberg1971,falk1977}. RSGs are also identified as the progenitor stars of SNe IIP SNe in direct detection efforts \citep{smartt2009b}. However, the masses of progenitor stars detected from pre-explosion images range from $8-15 M_{\odot}$ \citep{vandyk2003,li2005,li2006,maund2005a,maund2005b} which is closer to the lower mass end for a core-collapse event. No RSG star of mass $M > 17 M_{\odot}$ has been identified as a progenitor star of SNe IIP in direct detection efforts carried out in a volume limited sample \citep{smartt2009a}. In this context, a detailed census of type IIP SNe progenitors and the diversity of the properties of the SNe are important.

In a Type IIP SN, the fast moving stellar ejecta interacts with the circumstellar medium (CSM) created by the stellar wind of the progenitor star. The interaction creates a strong shock that moves ahead of the ejecta and is called the forward shock \citep{chevalier1982}. Electrons are accelerated to relativistic energies at the forward shock and emit at radio frequencies. Radio emission is absorbed at early times by either free-free absorption (FFA) or synchrotron self absorption (SSA) and can be modelled as synchrotron emission affected by either or both absorption processes. Depending on the dominant absorption process, various physical parameters  like mass-loss rate of the progenitor star, density of the CSM, ejecta density profile, shock deceleration parameter, magnetic field strength etc can be constrained from the modelled radio light curves and spectra \citep{chevalier1982,chevalier1998}. On the other hand, X-ray emission from Type IIP SNe can be either thermal or non-thermal in origin. Thermal X-rays can be emitted from the hot forward and/or reverse shock regions whereas non-thermal X-ray emission can be due to the Inverse Compton (IC) scattering. In Type IIP SNe, the plateau of the optical light curve is a phase of high density of ambient photons which can be IC scattered to X-ray energies by the relativistic electrons. As a result of IC scattering the relativistic electrons lose energy and there is a corresponding cooling break in the radio spectra above a characteristic frequency \citep{chevalier2006b}. 

In this work, we present long-term ($\sim$ 12 years) radio monitoring of a Type IIP, SN\,2004dj over a frequency range of 0.24 to 43 GHz. We model the radio light curves and spectra with the standard mini-shell model \citep{chevalier1982,chevalier1998} and derive the mass-loss rate of the progenitor star. We also search for the signatures of cooling in the radio spectra and interpret our results in light of other published results.

The paper is organised as follows: In \S \ref{sec:sn2004dj-literature-review}, we discuss the previous work published on SN\,2004dj. In \S \ref{sec:obs}, we present the GMRT and VLA observations of SN\,2004dj in detail. In \S \ref{sec:modelling}, we describe the standard model for radio emission. In \S \ref{mass-loss-rate calculation}, we calculate the mass-loss rate of the progenitor star from the modelled parameters. In \S \ref{sec:cooling}, we discuss the evolution of spectral indices and the signatures of cooling in the radio spectra. We summarise our main results in \S \ref{sec:summary}.  

\section{SN\,2004dj}
\label{sec:sn2004dj-literature-review}
SN\,2004dj was discovered by K. Itagaki \citep{nakano2004} in NGC\,2403 on 2004 July 31 (UT; all dates in this paper are in UT) with a visual magnitude of 11.2 mag at a position of $\alpha_{\rm J2000}$ = 07$^{\rm h}$37$^{\rm m}$17.02$^{\rm s}$, $\delta_{\rm J2000}$ = +65$^{\circ}$35$^{\prime}$57.8$^{\prime \prime}$. The SN was classified as a type IIP event from the optical spectrum taken on 2004 Aug 03.17 \citep{patat2004}. SN\,2004dj was discovered during the plateau phase i.e $\sim$ 1 month after the explosion and hence the maxima of the optical light curve was not observed.  

There is a large uncertainty in the explosion date of the SN. \cite{zhang2006} derived an explosion date of 2004 June 10 $\pm$ 21 assuming that SN\,2004dj evolves like SN\,1999em. \cite{chugai2005} derived an explosion date of 2004 June 13 assuming that the light curves of SN\, 2004dj and SN\,1999gi are similar. From the first one year of optical light curve of SN\,2004dj, \cite{vinko2006} infered an explosion date of 2004 June 30 $\pm$ 20 using Expanding Photosphere Method (EPM). \cite{chugai2007} used an explosion date of 2004 June 28 to fit the H$\alpha$ lines \citep{korcakova2005} observed on day 55 and 64 post-explosion. We assume the date of explosion as 2004 June 28 through out this paper since this is the latest among the series of papers that report the explosion dates and is consistent with the majority of the other explosion dates reported in the literature within error bars.

 \cite{patat2004} reported the photospheric expansion velocities of $H_{\alpha}$, $H_{\beta}$, Fe II (516.9 nm), Fe II (501.8 nm) and Fe II (492.4 nm) lines from the spectra obtained on 2004 Aug 3.17 as 6700, 5500, 4150, 3800 and 3850 km s$^{-1}$ respectively. \cite{vinko2006} reported the radial velocities of the SN ejecta from H$\alpha$ lines ranging from 6790 to 4256 km\,s$^{-1}$ on 33 to 99 days post-explosion, assuming the date of explosion as 2004 June 30 $\pm$ 20. \cite{chugai2007} derived the velocity from optical lines for SN\,2004dj on day $\sim$ 64 as 8200 km\,s$^{-1}$ assuming an explosion date of 2004 June 28.

The progenitor massive star of SN\,2004dj is identified to be a member of the compact star cluster Sandage 96 (S96) from the archival data \citep{maiz2004b}. For a cluster age of 13.6 Myr, \cite{maiz2004a} infered the mass of progenitor star as 15 $M_{\odot}$. \cite{wang2005} derived a cluster age of $\sim$ 20 Myr and the progenitor stellar mass as $\sim$ 12 $M_{\odot}$. \cite{vinko2006} estimated the progenitor mass by fitting the spectral energy distribution of S96 with theoretical models. While the solutions were similar to that of \cite{maiz2004a} and \cite{wang2005}, an additional solution resulted in a cluster age $\sim$ 8 Myr and progenitor mass of $M > 20 M_{\odot}$. It is interesting to note that, for any type IIP SNe, the highest progenitor mass detected from direct methods is 15 $M_{\odot}$ $<$ $M_{\rm ZAMS}$ $<$ 20 $M_{\odot}$ \citep[for SN\,1999ev;][]{maund2005a}.

SN\,2004dj was observed in different wave bands of electromagnetic spectrum like infrared \citep[IR;][]{kotak2005,sugerman2005}, X-ray \citep{pooley2004} and radio \citep{chandra2004,stockdale2004,beswick2005}. SN\,2004dj exploded at a distance of $D \sim$ 3.1 Mpc \citep{freedman2001} and is one of the nearest Type IIP event. \cite{vinko2006} infered a distance of $D$ = 3.47 $\pm$ 0.29 Mpc to the host galaxy NGC\,2403 from standard candle method and EPM. The distance to SN\,2004dj is taken as $D$ = 3.47 Mpc through out this paper. 

\subsection{Published radio and X-ray observations}
Radio emission was detected from SN\,2004dj at 8.4 GHz on 2004 Aug 02 with the Very Large Array \citep[VLA;][]{stockdale2004}. The first detection at 5 GHz was made with the  Multi-Element Radio Linked Interferometer Network (MERLIN) on 2004 Aug 5 with a flux density of 1.50 $\pm$ 0.15 mJy \citep{argo2004}. Radio emission was also detected at 1.4 GHz with the Giant Metrewave Radio Telescope \citep[GMRT;][]{chandra2004}. \cite{beswick2005} monitored SN\,2004dj with the MERLIN at 5 GHz starting from 2004 Aug 5 to Dec 2. At later epochs, the flux density of the SN was decreasing and hence the MERLIN observations placed a limit on the time of the peak of  5 GHz light curve \citep{beswick2005}. Assuming the first detection at 5 GHz on 2004 Aug 5 to be coincident with the peak of light curve, \cite{beswick2005} estimated the peak spectral luminosity as L$_{\rm 6 \, cm \, peak}$ = 2.45 $\times$10$^{25}$ erg s$^{-1}$ Hz$^{-1}$. This peak spectral luminosity is comparable to that of SN\,1999em i.e 2.2 $\times$ 10$^{25}$ erg s$^{-1}$ Hz$^{-1}$ \citep{pooley2002}. 

X-ray emission was detected from SN\,2004dj \citep{pooley2004} with the \textit{Chandra X-ray observatory} on 2004 Aug 9. \textit{Chandra} observations of SN\,2004dj was carried out at four epochs (2004 Aug 9, Aug 23, Oct 3 and Dec 22) and \cite{chakraborti2012} jointly fitted the X-ray spectra with the combination of a power law (IC component) and collisionaly ionized diffuse gas (thermal component). From thermal X-ray emission, they measured the mass-loss rate of the progenitor star as $\dot{M}$ = (3.2 $\pm$ 1.1) $\times$ 10$^{-7}$ $M_{\odot}$yr$^{-1}$. From the combination of radio, optical and X-ray data, \cite{chakraborti2012} derived the fraction of post shock energy density used in the amplification of magnetic fields to be $\epsilon_{B}$ = 0.082 and in accelerating electrons to relativistic energies to be $\epsilon_{e}$ = 0.39.    
\section{Observations and Data Analysis}
\label{sec:obs}

The VLA started observing SN\,2004dj from 2004 Aug 02.9 UT. The observations were carried out in the 8.46, 22.46 and 43.34 GHz bands. Radio emission was detected from the SN in the 8.46 GHz band \citep{stockdale2004}. Eventually, the SN was extensively observed in all VLA bands from 1.4 GHz (L-band) to 44 GHz band. J0642+679 and J0921+622 have been used as the phase calibrators to correct for the phase variations due to atmospheric fluctuations. 3C48 and 3C286 have been used as the primary calibrators to calibrate the flux density scale. 

The GMRT started observing SN\,2004dj from 2004 Aug 12 and detected radio emission from the SN at 1.4 GHz \citep{chandra2004}. Extensive follow-up observations were carried out with the GMRT at multiple epochs in the 1.4, 0.610, 0.325 and 0.235 GHz bands till 2016. 3C48 and 3C286 were used as the flux density calibrators, whereas J0835+555, J0614+607, J0614+617 and J0617+604 were used as the phase calibrator. 

Both the VLA and GMRT data were analyzed using standard Astronomical Image Processing software (AIPS) packages. Few data sets were not used due to the bad quality. We also included the 5 GHz MERLIN observations from \cite{beswick2005} for futher analysis. 

 \subsection{Radio lightcurves and spectral indices}

\begin{figure*}
\begin{centering}
\includegraphics[scale=0.65]{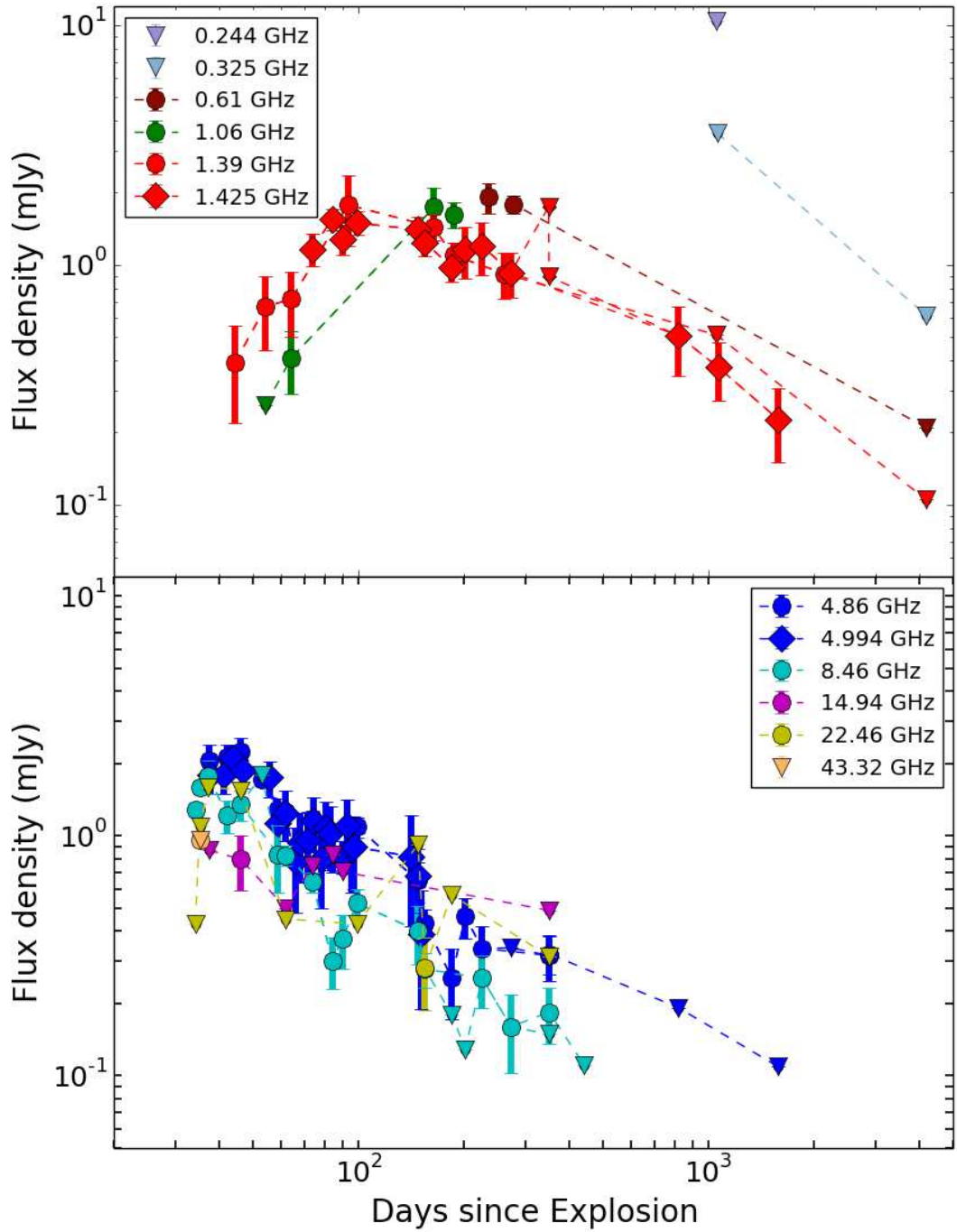}
 \caption{ \scriptsize{Upper panel: Light curves of SN 2004dj at frequencies 0.24, 0.33, 0.61, 1.06, 1.39 and 1.43 GHz. The 1.40 GHz data denoted in red color include both 1.39 GHz GMRT measurements (circles) and 1.43 GHz VLA measurements (diamonds). Lower panel: Light curves of SN 2004dj at frequencies 4.86, 4.99, 8.46, 14.94, 15.00, 22.46 and 43.32 GHz. Inverted triangles in both top and bottom panels denote 3$\sigma$ upper limits. The days since explosion are calculated assuming the date of explosion as 2004 June 28 (UT).}}
    \label{full-lc-subplot}
    \end{centering}
\end{figure*}

\begin{figure*}
    \includegraphics*[width=1.0\textwidth]{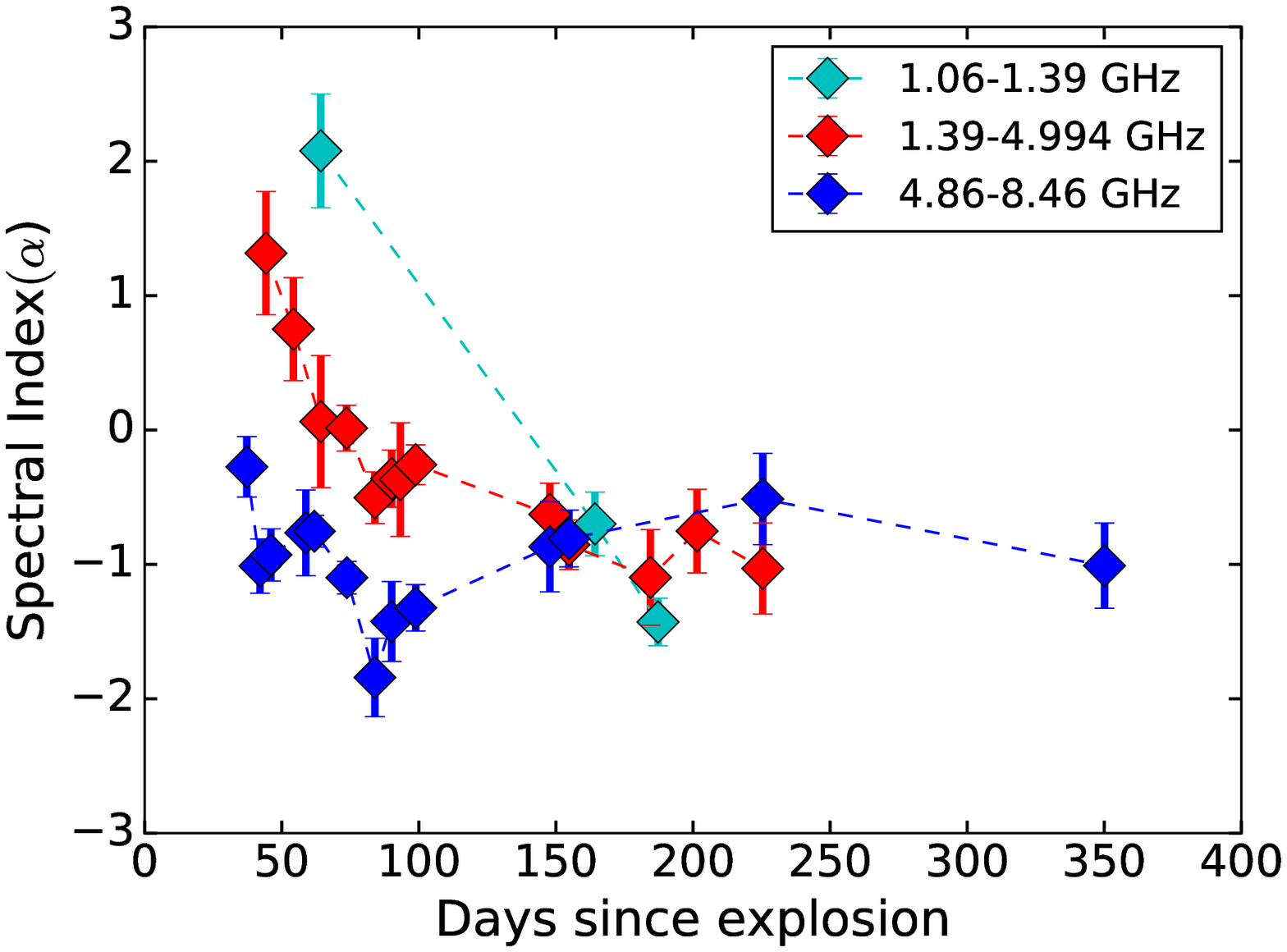}
\caption{ \scriptsize {The radio spectral index variation for SN\,2004dj. The figure shows the spectral index between frequencies 1.06/1.4, 1.4/4.9, 4.9/8.46 GHz. The days since explosion are calculated assuming the date of explosion as 2004 June 28 (UT). The spectral index curve between 4.9 GHz and 8.46 GHz shows a distinct dip around 75 days.}}
    \label{spectral-index variation}
\end{figure*}

We present the complete radio light curves at frequencies 0.24, 0.33, 0.61, 1.06, 1.4, 4.86, 1.994, 8.46, 14.94, 22.46 and 43.34 GHz in Figure \ref{full-lc-subplot}. All the light curves except the 1.4 and 1.06 GHz light curves are in the optically thin phase of the evolution. We also present the radio spectral index $\alpha$ ($F_{\nu} \propto \nu^{\alpha}$) using the flux density measurements at 1.06, 1.40, 5 and 8.46 GHz at various epochs in Figure \ref{spectral-index variation}. The temporal evolution of the spectral indices between 1.39 and 4.99 GHz traces the transition from optically thick to thin phase of the SN. The spectral indices between 4.86 and 8.46 GHz show values as steep as $-$ 1 and even lower for an extended period of time from $\sim$ day 50 to 150 post explosion. These values are steeper than the expected optically thin spectral indices of Type II SNe. We explain this trend in detail in terms of cooling in a later section (see \S \ref{first-sub-section-cooling}).

\begin{deluxetable*}{lccccccc}
\centering
\tablecaption{Details of radio observations of SN\,2004dj  \label{vla}}
\tablecolumns{6}
\tablenum{1}
\tablewidth{0pt}
\tablehead{
\colhead{Date of } & \colhead{Age\tablenotemark{a}} & \colhead{Frequency}  & \multicolumn2c{Supernova} &\multicolumn2c{Phase Calibrator} & \colhead{Telescope} \\
\cline{4-5} \cline{6-7}
\colhead{Observation (UT)} & \colhead{(day)} &
\colhead{(GHz)} & \colhead{F$_{\nu}$ (mJy)} & \colhead{rms (mJy\,beam$^{-1}$)} & \colhead{Name} & \colhead{F$_{\nu}$ (Jy)} & 
}
\startdata
	2004 Aug 1.88 & 34.12 & 8.46  & 1.28 $\pm$ 0.046 & 0.046 & J0642+679 & 0.32 & VLA-D Array     \\ 
	2004 Aug 1.90 & 34.14 & 22.46 & $<$ 0.429	& 0.169 & J0642+679 & 0.15 & VLA-D Array	   \\
2004 Aug 2.88 & 35.12 & 8.46	 &    1.6 $\pm$ 0.090 & 0.09  & J0921+622 & 0.78 & VLA-D Array     \\
2004 Aug 2.89 &  35.13 &   22.46	 & $<$1.095	& 0.416  & J0921+622 & 0.67 & VLA-D Array	\\
2004 Aug 2.91 &  35.15 &   43.32	 & $<$0.956	& 0.478 & J0921+622 & 0.55 & VLA-D Array	\\
2004 Aug 5.00 &  37.24   &  8.46  &  1.780 $\pm$ 0.289 &  0.228 & J0921+622 & 0.78 & VLA-D Array	\\
2004 Aug 5.01 &  37.25   &  22.46  & $<$1.593	 & 0.531 & J0921+622 & 0.61 & VLA-D Array	\\
2004 Aug 5.02 &  37.26  &  4.86  &  2.070 $\pm$ 0.322	& 0.247 & J0921+622 & 0.8 & VLA-D Array\\
2004 Aug 5.08 &  37.32  &  14.94	 & $<$0.870  & 0.290 & J0921+622	& 0.54 & VLA-D Array\\
2004 Aug 5.5 &  37.74  &  4.99 & 1.8 $\pm$ 0.3\tablenotemark{b}	 &   0.3 & -- & -- & MERLIN	\\
2004 Aug 8.8 & 41.04	  &  4.99 & 1.8	$\pm$ 0.3\tablenotemark{b} &  0.3	& -- & -- & MERLIN \\
2004 Aug 9.82&  42.06  &  8.46 &    1.217 $\pm$ 0.188 &  0.143 & J0921+622 & 0.78 & VLA-D Array \\
2004 Aug 9.83 & 42.07 &  4.86 &   2.133	$\pm$ 0.277 & 0.177 & J0921+622 & 0.8 & VLA-D Array	\\
2004 Aug 11.40	&  43.64  &  4.99 &  2.1 $\pm$ 0.3\tablenotemark{b} & 0.3 & -- & -- & MERLIN	\\
2004 Aug 12.24	&  44.24  &  1.39 &  0.39 $\pm$ 0.17 &  0.07 & J0614+607 & 0.97 & GMRT \\
2004 Aug 13.78	&  46.02  &  14.94	 &  0.799 $\pm$ 0.206 &  0.19 & J0921+622 & 0.61 & VLA-D Array	\\
2004 Aug 13.80 &  46.04  &  4.86 &  2.258 $\pm$ 0.299 & 0.196 & J0921+622 & 0.80 & VLA-D Array	\\
2004 Aug 13.80	&  46.04  &  8.46 & 1.349 $\pm$ 0.192 & 0.137 & J0921+622 & 0.78 & VLA-D Array	\\
2004 Aug 13.82	&  46.06  &  22.46 &  $<$1.551	&  0.517 & J0921+622 & 0.54 & VLA-D Array	\\
2004 Aug 14.50	&  46.74 &  4.99 &  1.88 $\pm$ 0.3\tablenotemark{b} &  0.3 & -- & -- & MERLIN	\\
2004 Aug 20.66	&  52.90 &  8.46 &  $<$1.8	&  0.644 & J0921+622 & 0.78 & VLA-D Array	\\
2004 Aug 20.67	&  52.91  &   4.86	&  1.72	$\pm$ 0.089  &  0.089 & J0921+622 & 0.8 & VLA-D Array	\\
2004 Aug 22.00	&  54.24  &  1.39 &    0.67	$\pm$ 0.23 & 0.10 & J0617+604 & 1.4 & GMRT\\
2004 Aug 22.00	&  54.24  &  1.06 &  $<$ 0.26 &  0.13 & J0614+617 & 1.19 & GMRT\\
2004 Aug 23.40	&  55.64  &  4.994 & 1.75 $\pm$ 0.3\tablenotemark{b}& 0.3  & -- & -- & MERLIN	\\
2004 Aug 26.54	&  58.78  &  8.46 &  0.839$\pm$ 0.262 & 0.065 & J0921+622 & 0.80 & VLA-D Array \\
2004 Aug 26.55	&  58.78 &  4.86 &  1.282 $\pm$ 0.083 &  0.083 & J0921+622 & 0.81 & VLA-D Array	\\
2004 Aug 26.60	&  58.84  &  4.99	&  1.125 $\pm$ 0.3\tablenotemark{b} & 0.3 & -- & -- & MERLIN	\\
2004 Aug 29.60	&  61.84  &  4.99 &  1.25 $\pm$ 0.3	& 0.3 & -- & -- & MERLIN	\\
2004 Aug 29.65	&  61.88  &  14.94	&  $<$0.5 & 0.158 & J0921+622 & 0.74 & VLA-D Array	\\
2004 Aug 29.66	&  61.89 & 4.86	&  1.26 $\pm$ 0.088	& 0.088	& J0921+622 & 0.8 & VLA-D Array \\
2004 Aug 29.67	&  61.90  & 8.46 &  0.83 $\pm$ 0.079 & 0.079 & J0921+622 & 0.78 & VLA-D Array	\\
2004 Aug 29.68	&  61.91 &  22.46 &  $<$0.450 &  0.149 & J0921+622 & 0.77 & VLA-D Array	\\
2004 Sep 01.00	&  64.24 &  1.39 & 0.72 $\pm$ 0.22	&  0.075 & J0614+607 & 1.23 & GMRT	\\
2004 Sep 01.00	&  64.24  &  1.06 &  0.41 $\pm$ 0.12 &  0.12 & J0614+607 & 1.42 & GMRT	\\
\enddata
\tablenotetext{a}{The age is calculated assuming date of explosion as 2004 June 28.00 (UT)}
\tablenotetext{b}{Beswick et al. (2005) and personal communication with Megan Argo.}
\end{deluxetable*}

\begin{deluxetable*}{lccccccc}
\centering
\tablecaption{Details of radio observations of SN\,2004dj (\textit{continuation})  \label{vla}}
\tablecolumns{6}
\tablenum{1}
\tablewidth{0pt}
\tablehead{
\colhead{Date of } & \colhead{Age\tablenotemark{a}} & \colhead{Frequency}  & \multicolumn2c{Supernova}& \multicolumn2c{Phase Calibrator} & \colhead{Telescope} \\
\cline{4-5} \cline{6-7}
\colhead{Observation (UT)} & \colhead{(day)} &
\colhead{(GHz)} & \colhead{F$_{\nu}$ (mJy)} & \colhead{rms (mJy\,beam$^{-1}$}) & \colhead{Name} & \colhead{F$_{\nu}$ (Jy)} & 
}
\startdata
2004 Sep 2.60 & 65.84  &  4.99 &  0.78 $\pm$ 0.3\tablenotemark{b} & 0.3 & -- & -- & MERLIN	\\
2004 Sep 4.90	&  68.14  &  4.99&   0.94 $\pm$ 0.3\tablenotemark{b} &  0.3 & -- & -- & MERLIN \\
2004 Sep 08.6 & 71.84  &  4.99 &  0.96 $\pm$ 0.3\tablenotemark{b}	& 0.3 & -- & -- & MERLIN	\\
2004 Sep 10.40	&  73.64 &  1.46 &  1.164 $\pm$ 0.180 & 0.138 & J0921+622 & 0.9 & VLA-A Array	\\
2004 Sep 10.50	&  73.74 &  4.88 &  1.184 $\pm$ 0.086 & 0.062 & J0921+622 & 0.76 & VLA-A Array	\\
2004 Sep 10.50	&  73.74  &  8.44 &  0.644 $\pm$ 0.062	&   0.05 & J0921+622 & 0.75 & VLA-A Array	\\
2004 Sep 10.56	&  73.80  &  14.94	&   $<$0.753 &  0.251 & J0921+622 & 0.74 & VLA-A Array	\\
2004 Sep 11.30	&  74.54  &  4.99  &  1.15 $\pm$ 0.3\tablenotemark{b} &  0.3 & -- & -- & MERLIN	\\
2004 Sep 14.70 &  77.94 &  4.99	&   0.8 $\pm$ 0.3\tablenotemark{b} &  0.3 & -- & -- & MERLIN	\\
2004 Sep 17.30	&  80.54  &  4.99 &  1.08 $\pm$ 0.3\tablenotemark{b} & 0.3 & -- & -- & MERLIN \\
2004 Sep 19.90	&  83.14  &  4.99 &  1.02 $\pm$ 0.3	&  0.3 & -- & -- & MERLIN	\\
2004 Sep 20.72	&  83.96  &  14.94	&  $<$0.835	&  0.3 & J0921+622 & 0.73 & VLA-A Array	\\
2004 Sep 20.73	&  83.97  &  4.86 &  0.838 $\pm$ 0.137 & 0.069 & J0921+622 & 0.73 & VLA-A Array	\\
2004 Sep 20.74	&  83.98  &  8.46 &  0.302 $\pm$ 0.073 & 0.051 & J0921+622 & 0.72 & VLA-A Array	\\
2004 Sep 20.75	&  83.99  &  1.425	&  1.555 $\pm$ 0.158 & 0.087 & J0921+622 & 0.88 & VLA-A Array	\\
2004 Sep 26.83	&  90.07  &  14.94	&  $<$0.708	&  0.236 & J0921+622 & 0.74 & VLA-A Array	\\
2004 Sep 26.84	&  90.08 & 4.86	 &  0.824 $\pm$ 0.128 & 0.07 & J0921+622 & 0.73 & VLA-A Array	\\
2004 Sep 26.85	&  90.09  &  8.46 &  0.374 $\pm$ 0.095 & 0.059 & J0921+622 & 0.72 & VLA-A Array \\
2004 Sep 26.86	&  90.10   &  1.425	 &  1.285 $\pm$ 0.187 & 0.098 & J0921+622 & 0.85 & VLA-A Array	\\
2004 Sep 29.40	&  92.64  &  4.99 &  1.11 $\pm$ 0.3	 &  0.3 & -- & -- & MERLIN	\\
2004 Sep 30.00	&  93.24  &  1.39 &  1.78 $\pm$ 0.58 & 0.23 & J0614+607 & 1.93 & GMRT	\\
2004 Oct 02.00	&  95.24   &  4.99  & 0.88 $\pm$ 0.3\tablenotemark{b}	&  0.3 & -- & -- & MERLIN	\\
2004 Oct 03.70	&  96.94  &  4.994	&  0.90 $\pm$ 0.3\tablenotemark{b} &   0.3 & -- & -- & MERLIN	\\
2004 Oct 5.60 &  98.84 &  1.425	&  1.498 $\pm$ 0.169 & 0.078 & J0921+622 & 0.90 & VLA-A Array	\\
2004 Oct 5.61	&  98.85  &  4.86	&    1.091 $\pm$ 0.104 & 0.089 & J0921+622 & 0.91 & VLA-A Array	\\
2004 Oct 5.62	&  98.86  &  8.46 &  0.524 $\pm$ 0.076 & 0.072 & J0921+622 & 0.8 & VLA-A Array	\\
2004 Oct 5.63 & 98.87  &  22.46	  &  $<$0.429&  0.143 & J0921+622 & 0.7 & VLA-A Array	\\
2004 Nov 16.50	&  140.74  &  4.99	&  0.820 $\pm$ 0.4\tablenotemark{b} &  0.4 & -- & -- & MERLIN	\\
2004 Nov 23.50	&  147.74  &  4.99 &    0.68 $\pm$ 0.2 & 0.2 & -- & -- & MERLIN	\\
2004 Nov 23.52	&  147.76  &  1.425	 & 1.402 $\pm$ 0.182 & 0.075 & J0921+622 & 0.86 & VLA-A Array	\\
2004 Nov 23.54	&  147.78  &   4.86	 &  0.649 $\pm$ 0.125 & 0.104 & J0921+622 & 0.69 & VLA-A Array	\\
2004 Nov 23.55	&  147.79  &  8.46	&  0.401 $\pm$ 0.11 & 0.109 & J0921+622 & 0.6 & VLA-A Array	\\
2004 Nov 23.55	&  147.79 &  22.46	&  $<$0.918	&  0.306 & J0921+622 & 0.31 & VLA-A Array	\\
2004 Nov 26.50	&  150.74  &  4.99	&  0.39 $\pm$ 0.2\tablenotemark{b} & 0.2 & -- & -- & MERLIN	  \\
2004 Nov 30.51	&  154.75  &  22.46	 &  0.282 $\pm$ 0.094 &  0.094 & J0921+622 & 0.72 & VLA-A Array	\\
\enddata
\tablenotetext{a}{The age is calculated assuming date of explosion as 2004 June 28.00 (UT).}
\tablenotetext{b}{Beswick et al. (2005) and personal communication with Megan Argo.}
\end{deluxetable*}

\begin{deluxetable*}{lccccccc}
\centering
\tablecaption{Details of radio observations of SN\,2004dj (\textit{continuation}) \label{vla}}
\tablecolumns{6}
\tablenum{1}
\tablewidth{0pt}
\tablehead{
\colhead{Date of } & \colhead{Age\tablenotemark{a}} & \colhead{Frequency}  & \multicolumn2c{Supernova} &\multicolumn2c{Phase Calibrator} & \colhead{Telescope} \\
\cline{4-5} \cline{6-7}
\colhead{Observation (UT)} & \colhead{(day)} &
\colhead{(GHz)} & \colhead{F$_{\nu}$ (mJy)} & \colhead{rms (mJy\,beam$^{-1}$}) & \colhead{Name} & \colhead{F$_{\nu}$ (Jy)} & 
}
\startdata
2004 Nov 30.51	&  154.75  &  1.425	 &  1.24 $\pm$ 0.154  &  0.091 & J0921+622 & 0.87 & VLA-A Array	\\
2004 Nov 30.52	&  154.76  &  4.86  &  0.435 $\pm$ 0.059 &  0.055 & J0921+622 & 0.77 & VLA-A Array	\\
2004 Nov 30.53	&  154.77  &  8.46  &  0.278 $\pm$ 0.045 & 0.043 & J0921+622 & 0.79 & VLA-A Array	\\
2004 Dec 09.00	& 163.24  &  1.39 &  1.44 $\pm$ 0.18 & 0.07 & J0617+604 & 1.25 & GMRT	\\
2004 Dec 10.00 &  164.24   &   1.06 &  1.74 $\pm$ 0.35 &  0.15 & J0614+607 & 0.3 & GMRT	\\
2004 Dec 30.21	&  184.45  &  1.425  &  0.979 $\pm$ 0.132 & 0.081 & J0921+622 & 0.86 & VLA-A Array	\\
2004 Dec 30.23	&  184.47  &  4.86 &  0.255 $\pm$ 0.084	 &  0.07 & J0921+622 & 0.75 & VLA-A Array	\\
2004 Dec 30.24	&  184.48 & 8.46 & $<$0.180	&  0.059 & J0921+622 & 0.80 & VLA-A Array	\\
2004 Dec 30.25	&  184.49 & 22.46 &  $<$0.567 &  0.189 & J0921+622 & 0.65 & VLA-A Array\\
2005 Jan 01.00	&  186.24  & 1.39 &  1.10 $\pm$ 0.14  & 0.05 & J0617+604 & 1.21 & GMRT	\\
2005 Jan 01.00	&  187.24 &  1.06 &  1.62 $\pm$ 0.20  &  0.14 & J0614+607 & 1.30 & GMRT	\\
2005 Jan 16.23	&  201.47 &  1.425	&  1.16 $\pm$ 0.281	& 0.164 & J0921+622 & 0.88 & VLA-BnA Array	\\
2005 Jan 16.24	&  201.48 & 4.86 &  0.461 $\pm$ 0.09 &   0.07 & J0921+622 & 0.81 & VLA-BnA Array	\\
2005 Jan 16.25	&  201.49  &  8.46	&  $<$0.129	&  0.045 & J0921+622 & 0.88 & VLA-BnA Array	\\
2005 Feb 8.19 &  225.43  &  1.425 &  1.2 $\pm$ 0.295 &  0.125 & J0921+622 & 0.86 & VLA-BnA Array \\
2005 Feb 8.20	&  225.44  &  4.86 &  0.339 $\pm$ 0.079	 &  0.065 & J0921+622 & 0.78 & VLA-BnA Array \\
2005 Feb 8.21&  225.45  &  8.46	 &  0.255 $\pm$ 0.063 & 0.044 & J0921+622 & 0.87 & VLA-BnA Array	\\
2005-Feb-18.00	&  235.24  &  0.61 &  1.92 $\pm$ 0.28 & 0.17 & J0834+555 & 8.16 & GMRT	\\
2005 Mar 15.00	&  260.24 & 1.39 &  0.92 $\pm$ 0.2 & 0.07 & J0614+607 & 1.21 & GMRT	\\
2005 Mar 26.96	&  272.20  &  1.425	 &  0.93 $\pm$ 0.2 & 0.125 & J0921+622 & 0.94 & VLA-B Array \\
2005 Mar 26.97	&  272.21  &  4.86 &  $<$0.341	&  0.07 & J0921+622 & 0.89 & VLA-B Array \\
2005 Mar 26.98	&  272.22  & 8.46 &  0.16 $\pm$ 0.058 & 0.036 & J0921+622 & 0.84 & VLA-B Array 	\\
2005 Apr 01.00	&  277.24  &  0.61	&  1.79	$\pm$ 0.16 & 0.243 & J0834+555 & 8.755 & GMRT	\\
2005 Jun 12.81	&  349.05  & 1.425 &  $<$1.750	&  0.487 & J0921+622 & 0.83 & VLA-CnB Array	\\
2005 Jun 12.82	&  349.06  &  4.86 &  0.317	$\pm$ 0.07 & 0.07 & J0921+622 & 0.81 & VLA-C Array \\
2005 Jun 12.84	&  349.08  &  8.44 & $<$0.149 &   0.050 & J0921+622 & 0.98 & VLA-C Array	\\
2005 Jun 13.78	&  350.02  &  22.485 &  $<$0.314 &  0.105 & J0921+622 & 0.79 & VLA-CnB Array	\\
2005 Jun 13.79	&  350.03   &  8.44  &   0.184 $\pm$ 0.048 & 0.047 & J0921+622 & 0.97 & VLA-C Array	\\
2005 Jun 13.82	&  350.06  &   14.94 &  $<$0.489 &  0.163 & J0921+622 & 0.97 & VLA-CnB Array	\\
2005 Jun 13.82	&  350.06  &   1.425  &  $<$0.896  &  0.30 & J0921+622 & 0.81 & VLA-CnB Array	\\
2005 Jun 13.84	&  350.08 &  4.885  &    0.322 $\pm$ 0.058 &  0.056 & J0921+622 & 0.83 & VLA-CnB Array	\\
2005 Sep 11.91	&  440.15  &  8.46	&  $<$0.111	&  0.037 & J0921+622 & 1.01 & VLA-CnB Array	\\
2006 Sep 25.54	&  818.78  &  1.425	 &  0.507 $\pm$ 0.162 &  0.162 & J0921+622 & 0.76 & VLA-CnB Array	\\
2006 Sep 25.55	&  818.79  &   4.86	 & $<$0.192	&  0.063 & J0921+622 & 1.1 & VLA-CnB Array	\\
\enddata
\tablenotetext{a}{The age is calculated using assuming date of explosion as 2004 June 28.00 (UT).}
\end{deluxetable*}

\begin{deluxetable*}{lccccccc}
\centering
\tablecaption{Details of radio observations of SN\,2004dj (\textit{continuation})  \label{vla}}
\tablecolumns{6}
\tablenum{1}
\tablewidth{0pt}
\tablehead{
\colhead{Date of } & \colhead{Age\tablenotemark{a}} & \colhead{Frequency}  & \multicolumn2c{Supernova} &\multicolumn2c{Phase Calibrator} & \colhead{Telescope} \\
\cline{4-5} \cline{6-7}
\colhead{Observation (UT)} & \colhead{(day)} &
\colhead{(GHz)} & \colhead{F$_{\nu}$ (mJy)} & \colhead{rms (mJy\,beam$^{-1}$}) & \colhead{Name} & \colhead{F$_{\nu}$ (Jy)} & 
}
\startdata
2007 May 15.24	&  1050.48  &   0.244 &  $<$10.47& 3.949 & J0834+555 & 8.70 & GMRT	\\
2007 May 18.41	&  1053.65 &  1.39 &  $<$0.510	&  0.169 & J0834+555 & 9.54 & GMRT	\\
2007 May 22.37	&  1057.61  &  0.325 &  $<$3.549 &   1.183 & J0834+555 & 9.77 & GMRT	\\
2007 May 28.84	&  1064.08  &  1.425 &  0.375 $\pm$ 0.102 &  0.058 & J0921+622 & 0.81 & VLA-A Array	\\
2008 Oct 24.48	&  1578.72    &  1.425	 &  0.228 $\pm$ 0.077  &  0.07 & J0921+622 & 1.07 & VLA-A Array	\\
2008 Oct 24.50	&  1578.74    &  4.86   &  $<$0.109 &  0.046 & J0921+622 & 1.50 & VLA-A Array \\
2016 Nov 02.04	&  4187.28    &  1.39   &  $<$0.105 &  0.035 & J0614+607 & -- & GMRT \\
2016 Oct 25.92	&  4180.16    &  0.610   &  $<$0.210 &  0.050 & J0834+555 & -- & GMRT \\
2016 Nov 05.04	&  4190.28    &  0.325   &  $<$0.615 &  0.095 & J0834+555 & -- & GMRT \\
\enddata
\tablenotetext{a}{The age is calculated assuming date of explosion as 2004 June 28.00 (UT).}
\end{deluxetable*}

\section{A Radio Model}
\label{sec:modelling}

We fit the data with the standard mini-shell model in which radio emission is purely synchrotron with either synchrotron self absorption (SSA) or free-free absorption (FFA) as the dominant absorption process \citep{chevalier1982, chevalier1998}. The model assumes that the magnetic field energy density and the energy density of relativistic electrons are proportianal to the post shock energy density \cite[model 1 of][]{chevalier1996}. The temporal and spectral evolution of radio flux density and optical depth can be modelled. The FFA model was proposed by \cite{chevalier1982} and was developed in detail by \cite{vandyk1994} and \cite{weiler2002} later. The radio flux density, $S(\nu,t)$, with FFA as the dominant absorption can be written as:
\begin{equation}
S(\nu,t)=K_{1} \left( \frac{\nu}{5\hspace{0.1 cm} \rm GHz}\right)^{\alpha} \left( \frac{t}{100\, \rm day}\right)^{\beta}e^{-\tau_{\rm ffa}}
\end{equation}

\begin{equation}
\tau_{\rm ffa}=K_{2}\left(\frac{\nu}{5\hspace{0.1 cm}\rm GHz}\right)^{-2.1}\left(\frac{t}{100\, \rm day}\right)^{\delta}
\end{equation}

where $\alpha$ is the optically thin spectral index $(F_{\nu}$ $\propto$ $\nu^{\alpha})$, which is related to the electron energy index p ($N (E) \propto$ $E^{-p}$ ) as $\alpha$=$-(p-1)/2$. Here $K_{1}$ and $K_{2}$ are the flux density and optical depth normalizations, respectively. $'t'$ denotes the days since explosion assuming a date of explosion 2004 June 28 \citep{chugai2007}. The parameter $\delta$ is related to the shock deceleration parameter m ($R \propto$ $t^{m}$) as $\delta$ = $-3m$. The parameter $\delta$ is essentially constrained by the evolution of the optical depth in time. If we look at the light curves (see Figure \ref{full-lc-subplot}) there are not enough simultaneous multi-frequency measurements in the optically thick part to allow the derivation of the optical depth behaviour in time. \cite{weiler1986} and  \cite{chevalier1984} suggest that in these cases it is recommented to do the fit with 4 parameters where $\delta$ = ($-$3 + $\alpha$ $-$ $\beta$). We fit the full data with the FFA model keeping $K_{1}$, $K_{2}$, $\alpha$ and $\beta$  as the free parameters.

We use the formulation of SSA model from \citep{chevalier1998}. The variation of radio flux density and SSA optical depth can be written as:

\begin{equation}
S(\nu,t)=K_{1} \left( \frac{\nu}{5\hspace{0.1 cm} \rm GHz}\right)^{5/2} \left( \frac{t}{100\hspace{0.1 cm} \rm day}\right)^{a} \left(1- e^{-\tau_{\rm ssa}} \right) 
\end{equation}

\begin{equation}
\tau_{\rm ssa}=K_{2}\left(\frac{\nu}{\rm 5\hspace{0.1 cm} GHz}\right)^{-(p+4)/2}\left(\frac{t}{100\hspace{0.1 cm}  \rm day}\right)^{-(a+b)}
\end{equation}

Here, $a$ and $b$ denote the temporal index of the radio flux density in the optically thick ($F\propto t^{a}$) and thin phases ($F\propto t^{-b}$), respectively. Assuming the energy density in the magnetic field and relativistic electrons are proportional to the post-shock energy density \citep[model 1 of][]{chevalier1996}, the shock deceleration parameter $m$ is related to $a$, $b$ and $p$ as $a=2m + 0.5$ in the optically thick phase and $b = (p + 5 -6m)/2$ in the optically thin phase \citep{chevalier1998}. 

We perform a two variable ($\nu$,$t$) fit to the complete data for both FFA and SSA models. The free parameters are $K_{1}$, $K_{2}$, $\alpha$ and $\beta$ in the FFA model and $K_{1}$, $K_{2}$, $m$ and $p$ in the SSA model. FFA and SSA model fits with the data with reduced chi-square values of 1.78 and 1.51 respectively. The best-fit parameters for both FFA and SSA model fits are given in Table \ref{fittedpara}. Figure 3 and 4 shows the model fits along with the observed flux density measurements. While the reduced chi-square value for SSA model is marginally better than that of FFA model, it is difficult to infer that either of the models fit the data better from visual inspection or from the reduced chi-square values. This is due to the sparse data available in the optically thick regime. We derive the shock deceleration parameter $m$ = 0.93 $\pm$ 0.02 and $m$ = 0.97 $\pm$ 0.01 for FFA and SSA models respectively.

 We also repeated the modelling with the date of explosion as 2004 June 10 \citep{zhang2006}. The reduced chi-square and physical parameters are not significantly different from the previous values. While the date of explosion of SN\,2004dj is uncertain due to the late discovery, our results are not critically sensitive to the exact date of explosion.

\subsection{Dominat absorption mechanism}

Since the FFA or SSA models cannot be clearly distinguished from the fits, we can look for other signatures that can disentangle the dominant absorption processes. Assuming SSA as the dominant absorption process that defines the peak flux density of the radio light curves, we can derive the size of the radio-emitting shell using the equation \citep{chevalier1998}: 

\begin{eqnarray}
R_{p} = 8.8 \times 10^{15} f_{\rm eB}^{-1/19} \left( \frac{f}{0.5} \right) ^{-1/19} \left( \frac{F_{p}}{\rm Jy} \right) ^{9/19} \nonumber \\* \times \left( \frac{D}{\rm Mpc} \right)^{18/19}  \left(\frac{\nu}{5\,\rm GHz} \right)^{-1} \rm cm
\end{eqnarray}

Here $f_{eB}$ is the equipartition factor defined as $f_{eB} = \epsilon_{e}/\epsilon_{B}$ where $\epsilon_{e}$ denotes the relativistic electron energy density and $\epsilon_{B}$ denotes the magnetic field energy density. $F_{p}$ is the peak flux density of the light curve at frequency $\nu$. $f$ is the volume filling factor of the spherical emission region of radius $R$. $D$ is the distance to the SN in units of Mpc. The equation for $R_{p}$ is specifically for electron energy index $p = 3$ \citep{chevalier1998}. However, we use this formula since our best fit $p$ value for SSA model is $p = 2.9$ which is very close to 3. The mean shell velocity $v_{p}$ = $R_{p}$/t where $t$ is the time at which the light curve peaks. If the observed mean shell velocity is larger than this value, it means that SSA flux is low and other mechanism like FFA could be dominant in determining the peak flux. For SN\,2004dj, the peak flux density at 5 GHz is $F_{p}$ = 1.8 mJy on 2004 Aug 5 \citep[$t_{p}$ = 38 days;][]{beswick2005}. We use $f_{eB}$ = 4.8 \citep{chakraborti2012} and $f$ = 0.5 \citep{chevalier1998} to estimate the mean shell velocity $v_{p}$ $\sim$ 4007 km\,s$^{-1}$ on $\sim$ 38 days post explosion. The actual mean velocity obtained from optical line measurements can be now compared to this value. The velocity derived from the $H_{\alpha}$ lines for SN\,2004dj on day $\sim$ 36 (2004 Aug 3.17) is 6700 km\,s$^{-1}$ \citep{patat2004} and is larger than the velocity deduced above. \cite{vinko2006} reports the radial velocities of the SN ejecta from H$\alpha$ lines as 6790 on 2004 Aug 3. \cite{chugai2007} reports the shell velocity to be $\sim$ 8200 km\,s$^{-1}$ on 64 days post explosion from $H_{\alpha}$ lines. Thus these observations are suggestive of FFA as the dominant absorption process. \cite{chevalier2006b} also suggest FFA as the dominant absorption process for SN\,2004dj from the steep rise of 1.4 GHz light curve \citep{chandra2004}. We repeat this exercise using the 1.4 GHz light curve that peaks around day 93. The peak flux density is $F_{p}$ = 1.78 mJy at 1.4 GHz and the deduced mean velocity is $v_{p}$ $\sim$ 5830 km\,s$^{-1}$. The observed $H_{\alpha}$ line velocity on day 96 post explosion is 4373 km\,s$^{-1}$ \citep{vinko2006}. However, \cite{chugai2007} reports the observed shell velocity on day 98 post explosion to be $\sim$ 8000 km\,s$^{-1}$ from a prominent notch like feature in the spectra. This is also greater than the derived mean shell velocity and is indicative of FFA as the plausible absorption process.

\begin{deluxetable}{cc}
\centering
\tablecaption{Best fit parameters for FFA and SSA models \label{fittedpara}}
\tablecolumns{2}
\tablenum{2}
\tablewidth{0pt}
\tablehead{
\colhead{FFA} &  \colhead{SSA}  \\ 
}
\startdata
$K_{1}$ = 0.82 $\pm$ 0.02 &   $K_{1}$ = 62.46 $\pm$ 07.01  \\
$K_{2}$ = (3.01 $\pm$ 0.35) $\times$ 10$^{-2}$ &  $K_{2}$ = (1.27 $\pm$ 0.15) $\times$ 10$^{-2}$  \\
$\alpha$ = $-$0.92 $\pm$ 0.05 & $m$ = 0.97 $\pm$ 0.01  \\
$\beta$ = $-$1.13 $\pm$ 0.06 & $p$ = 2.87 $\pm$ 0.10   \\
$\chi_{\mu}^{2}$ = 1.78 & $\chi_{\mu}^{2}$ = 1.51  \\
\enddata
\tablecomments{$K_{1}$ and $K_{2}$ are the normalization parameters of flux density and optical depth, respectively. In FFA model, $\alpha$ and $\beta$ denotes the spectral and temporal evolution of the radio flux density. In the SSA model m denotes the shock deceleration parameter and p denotes the electron energy index.}
\end{deluxetable}

\begin{figure*}
\label{lc-fit}
    \includegraphics*[width=1\textwidth]{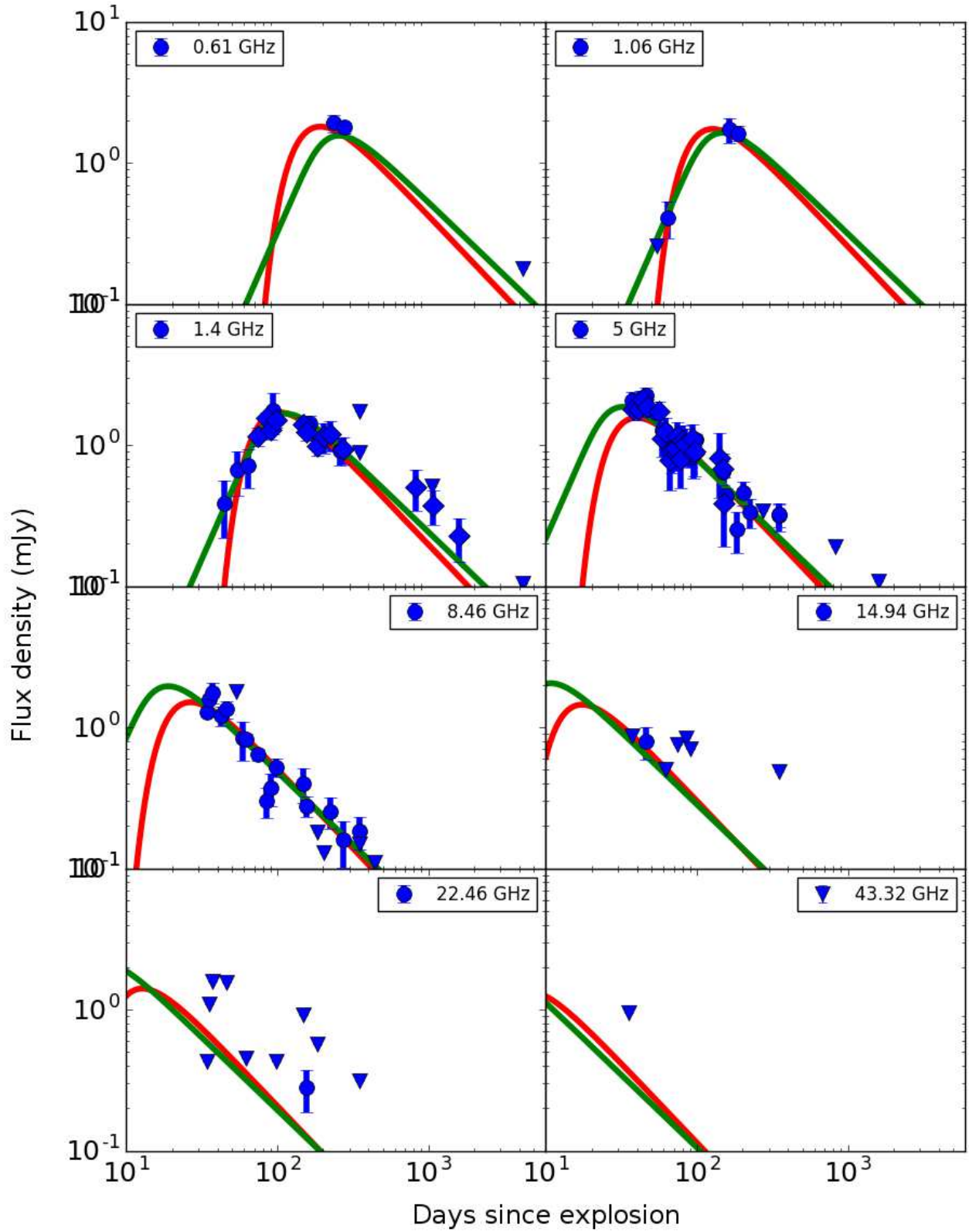}
\caption{ \scriptsize{ SSA and FFA model fit to the radio light curves of SN 2004dj at 0.61, 1.06, 1.4 , 5.00, 8.46, 14.94, 22.46 and 43.34 GHz bands. The 1.40 GHz data includes both 1.39 GHz GMRT measurements (circles) and 1.43 GHz VLA measurements (diamonds). Green solid line denotes the SSA model and red solid line denotes the FFA model. The days since explosion is calculated assuming the date of explosion as 2004 June 28 (UT).}}
\end{figure*}

\begin{figure*}
\label{spectra-fit}
    \includegraphics*[width=1\textwidth]{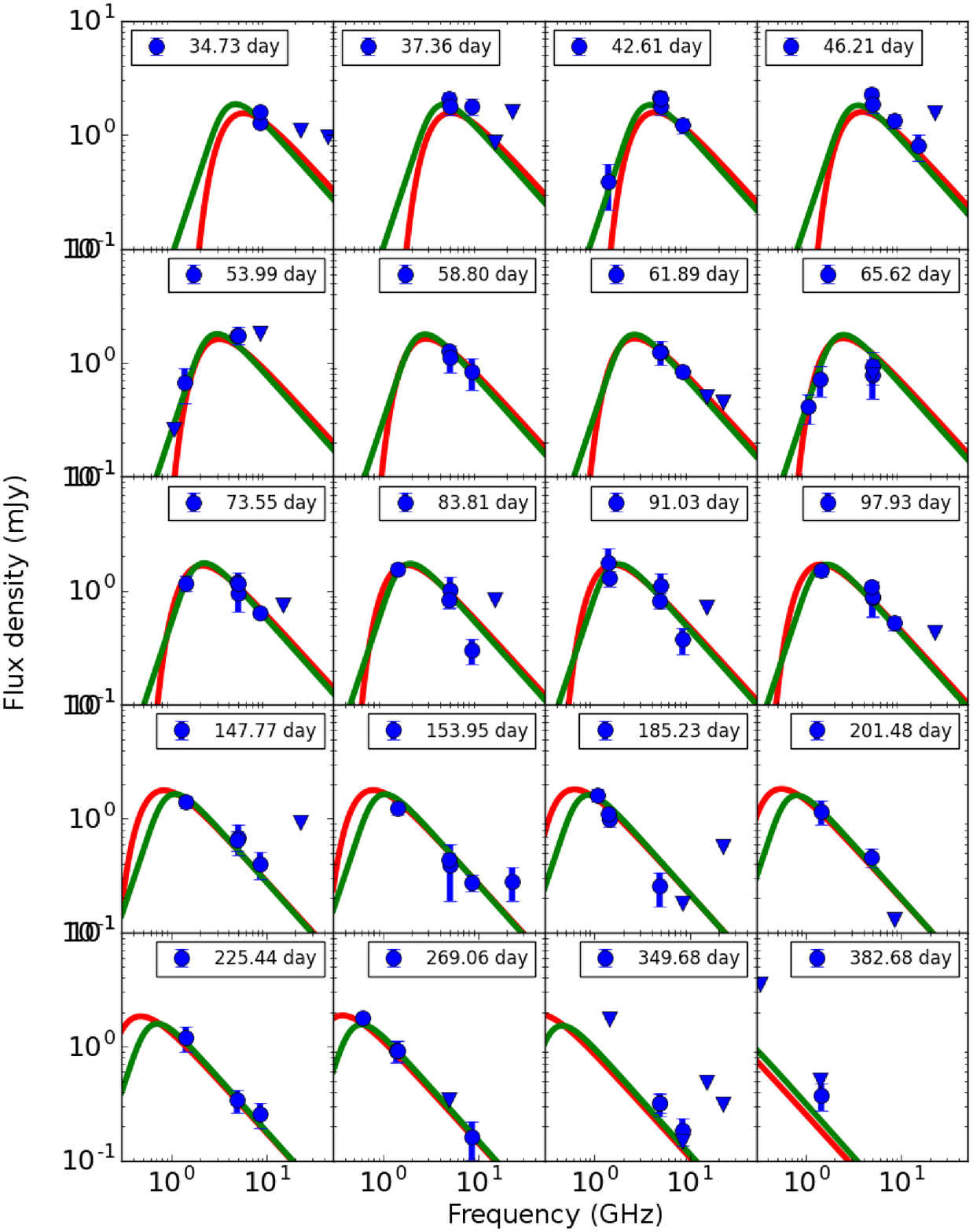}
\caption{\scriptsize{SSA and FFA model fit to the radio spectra of SN 2004dj on $\sim$ day 35 to 383 post explosion. Green solid line denotes the SSA model and red solid line denotes the FFA model. The days since explosion is calculated assuming the date of explosion as 2004 June 28 (UT).}}
\end{figure*}
 
\section{Mass-loss rate}
\label{mass-loss-rate calculation}

Assuming FFA as the dominant absorption process, we derive the mass-loss rate of the progenitor star of SN\,2004dj. The various assumptions that go into the calculation are as follows and are detailed in \cite{weiler1986}. The progenitor expels the outer layer of stellar material via constant stellar wind creating a density field around the star such that $\rho_{\rm wind} \sim r^{-2}$. The wind material is completely ionized possibly due to the initial flash of radiation from the SN. Assuming electron ion equilibrium and the wind material to be singly ionized with cosmic abundance, the mass loss rate is given by \cite{weiler1986} as

\begin{eqnarray}
\dot{M} = 3.02 \times 10^{-6} \tau^{0.5}_{5 \, \rm GHz} \left(  \frac{w}{10 \, \rm km\, s^{-1}} \right) \left( \frac{v_{i}}{10^{4}\, \rm km\,s^{-1}} \right)^{1.5} \nonumber \\* \times \left( \frac{t_{i}}{45\, \rm days}\right)^{1.5} \left(\frac{t}{t_{i}} \right)^{1.5\,m} m^{-1.5} \nonumber \\ * \times \left( \frac{T_{e}}{10^{4}\, \rm K} \right) ^{0.68} M_{\odot} \rm yr^{-1}   
\end{eqnarray}

where $w$ is the velocity of the stellar wind in km\,s$^{-1}$, $v_{i}$ is the SN ejecta velocity in km\,s$^{-1}$ at time t$_{i}$ (days post explosion) inferred from optical line observations. $T_{e}$ is the electron temperature in the stellar wind and is an uncertain parameter. We use $v_{i}$ = 8200 km\,s$^{-1}$, the highest velocity $H\alpha$ absorption feature seen in the optical spectra on $t_{i}$ = 64 days \citep{chugai2007}. From the modelled radio light curve (equation 2), we have $\tau_{5\, \rm GHz}$ (day 100) = 0.03 and m= 0.93. We derive the mass-loss rate of the progenitor star to be $\dot{M}/w1 = (1.37 \pm 0.11)\,\,T_{cs4}^{0.68}$ $\times$ $10^{-6}$ $M_{\odot}\rm yr^{-1}$. Here $w1$ denotes the stellar wind velocity in units of 10 km\,s$^{-1}$ and $T_{cs4}$ denotes the CSM electron temperature in units of $10^{4}$ K.

\section{Discussion}
\label{sec:cooling}

\subsection{Progenitor properties}
We model the radio observations with the standard mini-shell model \citep{chevalier1982} as explained in \S \ref{sec:modelling}. We estimate the shock deceleration parameter $m$ $\sim$ 0.9 (where $R \propto t^m$) for FFA model indicative of a mildly decelerating blast wave. This is in accord with the physics of the process as the blast wave is interacting with the CSM and is expected to slow down. The range of m values for Type II SNe is $m = 0.8 - 1$ \citep{weiler1986}. For a shocked shell of radius $R$ $\sim$ $t^{0.9}$, the ejecta density index n ($\rho \sim r^{-n}$) is given by $m = (n-3)/(n-2)$, assuming that the CSM is created by a steady stellar wind of density ($\rho \sim r^{-2}$) \citep{chevalier1982}. Thus for $m$ $\sim$ 0.9, we derive the power law ejecta density as $\rho \sim r^{-11.4}$. The progenitors of Type IIP SNe are understood to be RSGs with most of its hydrogen envelope intact during the SN explosion \citep{smartt2009b}. The value of $n$ for such a star is expected to be in the range $n = 7-12$ \citep{chevalier1982}. Thus the $n$ value derived from our analysis is consistent with a RSG progenitor of SN\,2004dj.

Assuming FFA as the dominant absorption process, we derive the mass-loss rate of the progenitor star of SN\,2004dj as $\dot{M} = (1.37 \pm 0.11) \times 10^{-6}\, M_{\odot}\rm yr^{-1}$. The progenitors of Type IIP SNe are understood to be red-supergiants whose initial masses range from 8 - 25 $M_{\odot}$ \citep{heger2003}. For the best studied RSGs, the range of mass-loss rates is considerably large, i.e  2 $\times$ 10$^{-7}$ to 1.5 $\times$ 10$^{-5}$ $M_{\odot}\rm yr^{-1}$ \citep{jura1990,van2005}. The mass-loss rates of RSGs at the time of explosion is estimated using stellar evolutionary calculations as $\sim$ 3 $\times$ 10$^{-7}$ to 3 $\times$ 10$^{-5}$ $M_{\odot}\rm yr^{-1}$ \citep{schaller1992,chevalier2006b}. The theoretical estimate of mass-loss rate for 15$M_{\odot}$ models is derived as $\dot{M}$ = (0.84-1.6) $\times$ 10$^{-6}$ $M_{\odot}\rm yr^{-1}$ and for 20$M_{\odot}$ models is derived as $\dot{M}$ = (3.0-6.2) $\times$ 10$^{-6}$ $M_{\odot}\rm yr^{-1}$ \citep{chevalier2006b}. Thus the mass-loss rate derived for SN\,2004dj from radio observations are consistent with the theoretical predictions for a progenitor star of mass $\sim$ 15$M_{\odot}$. The progenitor mass of SN\,2004dj was infered as 15$M_{\odot}$ from stellar population studies of S96 cluster \citep{maiz2004a}.

Mass-loss rate of the progenitor star of SN\,2004dj was derived by \cite{chakraborti2012} from X-ray emission measure as $\dot{M}$ = (3.2 $\pm$ 1.1) $\times$ 10$^{-7}$ $M_{\odot}\rm yr^{-1}$ which is $\sim$ 4 times lower than our mass-loss estimate. \cite{chugai2007} derived the mass-loss rate of SN\,2004dj to be $\sim$ 1 $\times$ 10$^{-6}$ $M_{\odot}\rm yr^{-1}$ for a wind velocity of 10 km\,s$^{-1}$, consistent with the mass-loss rate estimate from our analysis. \cite{chevalier2006b} derives a mass-loss rate of $\dot{M}_{-6}/w1$ $\sim$ (2-3) $T_{\rm cs5}^{3/4}$ for SN\,2004dj using the first 100 days of radio data. Here $\dot{M}_{-6}$ is the mass-loss rate in units of $10^{-6} M_{\odot}\rm yr^{-1}$ and $w1$ is the stellar wind velocity in units of 10 km\,s$^{-1}$. $T_{\rm cs5}$ denotes the CSM electron temperature in units of $10^{5}$ K. The authors compile the radio and X-ray data of all type IIP SNe available then and derive a range of mass-loss rates (1-10) $\times$ 10$^{-6}$ $M_{\odot}\rm yr^{-1}$ (see Table \ref{comparison-typeIIp}) for Type IIP progenitors and our mass-loss rate estimation is consistent with this range. 

\subsection{Signatures of cooling}
\label{first-sub-section-cooling}

The radio light curve and spectra of Type IIP SNe can be affected by cooling. Cooling becomes important depending on various parameters of SN such as ejecta, magnetic field strength, circumstellar medium, relativistic electrons etc \citep{chevalier2006b}. The electron can lose energy by adiabatic expansion, synchrotron cooling, and IC cooling. The dominant cooling process can be identified by calculating the break frequencies and cooling time scales for different cooling processes.  

One of the important signature of cooling is imprinted in the evolution of radio spectral indices. As a result of cooling the spectral index $\alpha$ steepens by $\Delta \alpha$ $\sim$ $-$0.5. This appears as a break in the spectra at a certain frequency when the electrons radiating above that characteristic frequency loses significant energy. 

Assuming that the synchrotron loss time scale is equal to the age of the SN, the expression for synchrotron break frequency is given by \cite{chevalier2006b} as
\begin{eqnarray}
\nu_{\rm syn} = 240 \left( \frac{\epsilon_{B}}{0.1} \right)^{(-3/2)} \left( \frac{\dot{M}}{10^{-6} M_{\odot} \rm yr^{-1}} \right)^{-3/2} \nonumber \\* \times \left( \frac{w}{10\, \rm km\,s^{-1}}\right)^{3/2} \left( \frac{t}{60\, \rm days} \right) \rm GHz   
\end{eqnarray}

Assuming that the IC cooling time scale is equal to the age of the SN, the IC break frequency is \citep{chevalier2006b}

\begin{eqnarray}
\nu_{\rm IC} = 8 \left( \frac{\epsilon_{B}}{0.1} \right)^{(1/2)} \left( \frac{\dot{M}}{10^{-6} M_{\odot} \rm yr^{-1}} \right)^{1/2} \nonumber \\* \times \left( \frac{v_{w}}{10\, \rm km\,s^{-1}}\right)^{-1/2} \left( \frac{t}{60\, \rm days} \right) \nonumber \\*
\left( \frac{V_{s}}{10^{4} \rm km\,s^{-1}}\right)^{4} \left( \frac{L_{\rm bol}(t)}{10^{42}\rm erg\,s^{-1}}\right) \rm GHz    
\end{eqnarray}
\cite{chakraborti2012} derived $\epsilon_{B}$ = 0.082 and $\epsilon_{e}$ = 0.39 for SN\,2004dj using four epochs of \textit{Chandra} data ( $\sim$ 42, 56, 97 and 177 days post explosion) where the authors found prominent IC X-ray component in the first two epochs. We calculate the break frequencies corresponding to synchrotron cooling and IC cooling using the above equations at these two epochs (i.e day 42 and 56 post explosion). The value of $\epsilon_{B}$ is taken from \citep{chakraborti2012} and $\dot{M}$ from our calculations. We use the value of shock velocity $V_{s}$ = 9.2 $\times$ $10^{3}$ km\,s$^{-1}$ \citep{chakraborti2012}. The bolometric luminosity during the plateau is $L_{\rm bol}$ $\sim$ 0.89 $\times$ 10$^{42}$ ergs \citep{zhang2006}. Asuming a wind velocity of 10 km\,s$^{-1}$, and temperature of $10^{4}$ K, the $\nu_{\rm syn}$ and $\nu_{\rm IC}$ corresponding to day 42 and 56 days post explosion are $\sim$ 141, 4 GHz and 188, 5 GHz  respectively. The synchrotron break frequency is too high to observe with the VLA during our observation epochs. The IC cooling break frequency is within our observation frequencies and  with the multi-frequency observations of SN\,2004dj, we can look for this signature in the data. In Figure \ref{spectral-index variation},  We plot the evolution of spectral indices between successive frequencies, 1.06/1.39, 1.39/4.99, 4.99/8.46 GHz with time for SN\,2004dj. The spectral index values of 4.86/8.46 GHz approaches values $\sim$ $-$1 and lower during an extended period starting from $\sim$ day 50 (see Figure \ref{spectral-index variation}). Thus we see the IC cooling break at $\sim$ 5 GHz, roughly consistent with the above calculation. The optical bolometric light curve is in the plateau phase during the same period \citep{zhang2006}. The dense optical photon medium in the plateau phase of the SN provides seed photons and enhances the IC cooling. This kind of a dip in the spectral index is seen for the second time in a type IIP supernova after SN\,2012aw \citep{yadav2014}. 

We establish IC cooling as the dominant cooling process as there is evidence of cooling break at $\sim$ 5 GHz in the radio spectral evolution. Here, we calculate the cooling timescale at $\sim$ 5 GHz for IC and synchrotron cooling process to further investigate this. 

The ratio of synchrotron cooling time scale to the adiabatic expansion timescale is \citep{chevalier2006b}

\begin{eqnarray}
\frac{t_{\rm syn}}{t} \approx 2.0 \left( \frac{\epsilon_{B}}{0.1} \right)^{-3/4} \left( \frac{\dot{M}_{-6}}{v_{w1}}\right) ^{-3/4} \left(\frac{\nu}{10\, \rm GHz} \right)^{-1/2} \nonumber \\* \times \left( \frac{t}{10\, \rm days} \right)^{1/2}  
\end{eqnarray}

The ratio of Compton cooling timescale to the adiabatic expansion timescale is 

\begin{eqnarray}
\frac{t_{\rm Comp}}{t} \approx 0.18 \left( \frac{L_{bol}}{2 \times 10^{42} \, \rm erg\, s^{-1}} \right) ^{-1} \left( \frac{\epsilon_{B}}{0.1} \right)^{1/4} \left( \frac{\dot{M}_{-6}}{v_{w1}}\right) ^{1/4} \nonumber \\* \times V_{s4}^{2} \left(\frac{\nu}{10\, \rm GHz} \right)^{-1/2}  \left( \frac{t}{10\, \rm days} \right)^{1/2}  
\end{eqnarray}
  
In general, lower values of $\epsilon_{B}$ and $\dot{M}$ favours IC cooling and especially at early times. We estimate the ratios $t_{\rm Comp}/t$ and $t_{\rm syn}/t$ at $\sim$ 42 and 56 days post explosion using the above equations as $\sim$ 0.9, 5.3  and 1.1, 6.1 respectively. While the ratios of cooling time-scale is close to 1 for IC cooling, the ratios are much higher for synchrotron cooling; clearly indicates that synchrotron cooling is not plausible. Thus these numbers favour the IC cooling to be in operation during these epochs at the SN shock. This is in agreement with the detection of non-thermal IC component in the X-ray spectra of SN\,2004dj during the first two epochs of observations ($\sim$ day 42 and 56 post explosion) by \cite{chakraborti2012}. Thus both cooling time scale and break frequency calculations supports the IC cooling happening at the SN shock during the plateau phase.

\begin{deluxetable*}{cccccccc}
\centering
\tablecaption{Comparison of SN\,2004dj parameters with other radio/X-ray bright Type IIP SNe. \label{comparison-typeIIp}}
\tablecolumns{8}
\tablenum{3}
\tablewidth{0pt} 
\tablehead{
\colhead{SN} &  \colhead{Parent} & \colhead{Distance} & $L_{\rm radio}$ & $L_{\rm X ray}$ & $\dot{M}$ & Progenitor mass & References \\ 
\colhead{-} &  \colhead{Galaxy} & \colhead{(Mpc)} & (erg s$^{-1}$ Hz$^{-1}$) & (erg s$^{-1}$) & ($10^{-6} M_{\odot}\rm yr^{-1}$) & ($M_{\odot}$) &  \\
}
\startdata
 SN\,1999em & NGC\,1637 & 11.7 $\pm$ 1.0 & 2.2 $\times$ 10$^{25}$ & 9 $\times$ 10$^{37}$ & 0.9\tablenotemark{a} & $<$15 & 1,2,3,4 \\
 SN\,1999gi & NGC\,3184  & 11.1$^{+2.0}_{-1.8}$ & --\tablenotemark{*} & 1.6 $\times$ 10$^{37}$ & $\sim$ 1\tablenotemark{b} & $<$12 & 1,2,3,5 \\
 SN\,2002hh & NGC\,6946 & 5.5 $\pm$ 1.0 &  1.3 $\times$ 10$^{25}$& 4 $\times$ 10$^{38}$ & 1.2\tablenotemark{a} & -- & 3,6,7,8  \\
 {\bf SN\,2004dj} & {\bf NGC\,2403} & {\bf 3.47} & {\bf 2.5 $\times$ 10$^{25}$} & {\bf 1.5 $\times$ 10$^{38}$} & {\bf 1.4\tablenotemark{c}}, (0.2 - 0.5)\tablenotemark{a}, 0.3\tablenotemark{b} & {\bf 15, $\sim$ 12, $>$20 } & 3,9,10,11   \\
 SN\,2004et & NGC\,6946 & 5.5 $\pm$ 1.0 & 8.7 $\times$ 10$^{25}$ & (2.1 - 3)$\times$ 10$^{38}$ & (1.6-1.8)\tablenotemark{a}, $\sim$ 2\tablenotemark{b}  & 15$^{+5}_{-2}$  & 3,6,12,13 \\
 SN\,2011ja & NGC\,4945 & 3.36 $\pm$ 0.09 & 7.3 $\times$ 10$^{24}$ & (1.3 - 5.5) $\times$ 10$^{38}$ & 0.12 - 1.9\tablenotemark{b}  & -- & 14,15 \\
 SN\,2013ej & M74 & 9.6 $\pm$ 0.7 & --\tablenotemark{*} & $\sim$ (1 - 4) $\times$ 10$^{42}$  & 2.6\tablenotemark{b} & -- & 16,17  \\
 SN\,2012aw & M95 & 10 & 7.2 $\times$ 10$^{25}$ & 9.2 $\times$ 10$^{38}$ & $\sim$ 3.2-10\tablenotemark{b} & 14-26, 17-18 & 18,19,20,21,22  \\
 SN\,2017eaw & NGC\,6946 & 5.86 $\pm$ 0.76 & 1.7 $\times$ 10$^{25}$ & 1.1 $\times$ 10$^{39}$ & -- & -- & 22,23,24  \\
\enddata
\tablecomments{This table is adapted from a similar table published by \cite{chevalier2006b} with more SNe added.}
\tablecomments{References: (1) \cite{leonard2002}, (2) \cite{smartt2003}, (3) \cite{chevalier2006b}, (4) \cite{pooley2002}, (5) \cite{schlegel2001}, (6) \cite{li2005}, (7) \cite{pooley2002iauc}, (8) \cite{beswick2005iauc}, (9) \cite{chakraborti2012}, (10) \cite{maiz2004a}, (11) \cite{vinko2006}, (12) \cite{misra2007}, \cite{li2005}, (13) \cite{argo2005}, (14) \cite{chakraborti2013}, (15) \cite{mouhcine2005}, (16) \cite{chakraborti2016} (17) \cite{bose2015}, (18)  \cite{kochanek2012}, (19) \cite{immler2012}, (20) \cite{fraser2012}, (21) \cite{vandyk2012}, (22) \cite{argo2017}, (23) \cite{bose2014}, (24) \cite{gref2017}.}
\tablecomments{X-ray luminosity in the energy range of 0.5 - 8 KeV for all SNe except SN\,2011ja, SN\,2017eaw (0.3 - 10 KeV) and SN\,2012aw (0.2 - 10 KeV). Radio luminosity is spectral luminosity at $\sim$ 5 GHz. Where possible, the X-ray and radio luminosities are taken at the peak of the light curve and for others, the limits of peak flux density is used. The mass of the progenitor stars are from the pre-explosion observations of the SN site. The listed mass-loss rates are from X-ray/radio observations and modelling.}
\tablenotetext{a}{The mass-loss rate is estimated from radio analysis and modelling assuming a stellar wind velocity of 10 km\,s$^{-1}$. While for SN\,1999em, SN\,2002hh, SN\,2004et \cite{chevalier2006b} and for SN\,2012aw \cite{yadav2014} calculate mass-loss rates assuming a CSM electron temperature of $10^{5}$ K, we recalculate these numbers for an electron temperature of $10^{4}$ K for comparison since we used electron temperature of $10^{4}$ K to calculate the mass-loss rate of SN\,2004dj.}
\tablenotetext{*}{Not detected in radio.}
\tablenotetext{b}{The mass-loss rates are estimated from the X-ray observations assuming a stellar wind velocity of 10 km\,s$^{-1}$.}
\tablenotetext{c}{The mass-loss rate estimated for SN\,2004dj from this work assuming a stellar wind velocity of 10 km\,s$^{-1}$ and CSM electron temperature of $10^{4}$ K.}
\end{deluxetable*}

\cite{chevalier2006b} predicts a flattened light curve and a dip in the radio light curve of Type IIP SNe especially at higher frequencies as a signature of IC cooling. However, from our error bars on flux densities and cadence of observation, it is difficult to look for a flattening or small dip in the light curve as predicted by \cite{chevalier2006b}. Cooling processes and its effect on the radio light curves could be modelled better with a good quality and high cadence early time data at radio frequencies. 

\begin{figure*}
\label{test}
    \includegraphics*[width=1\textwidth]{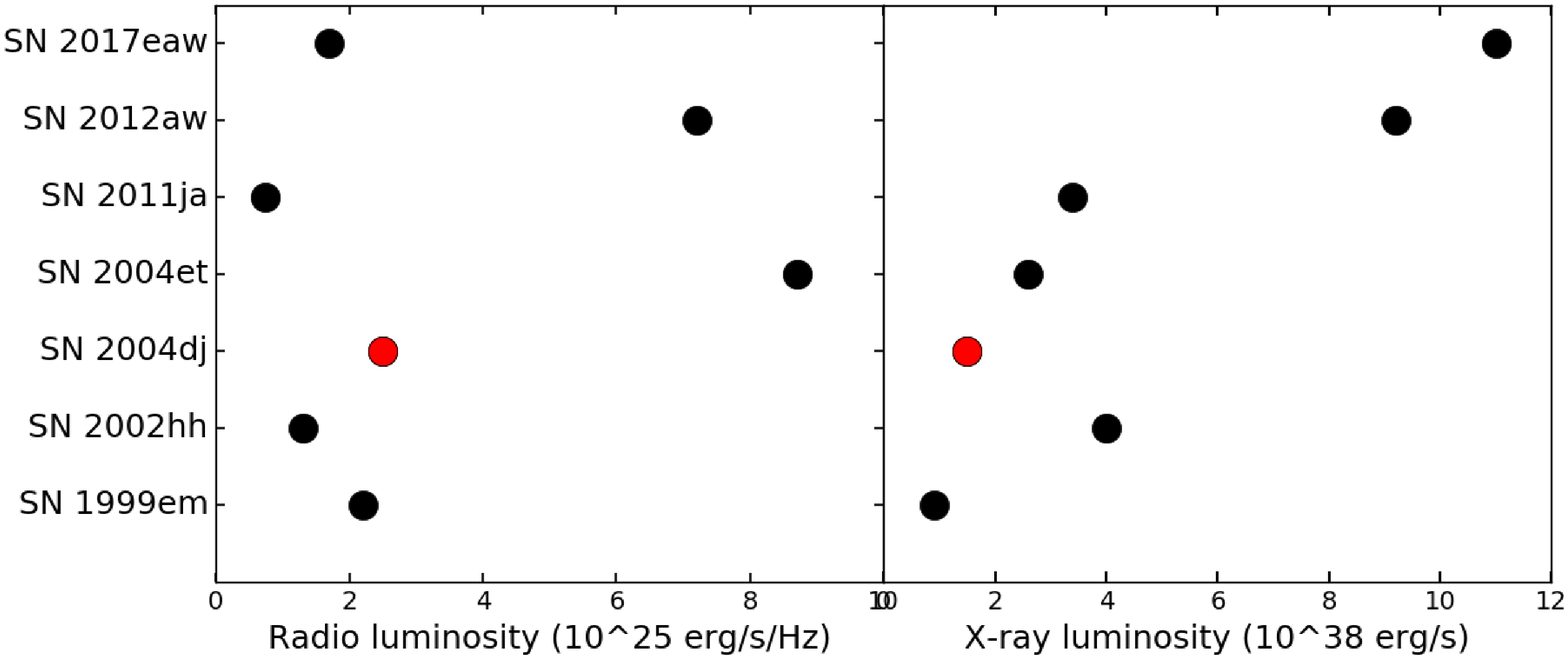}
\caption{ \scriptsize{Left panel:Radio spectral luminosity at $\sim$ 5 GHz of Type IIP SNe Right panel: X-ray luminosity in the energy range of 0.5 - 8 KeV for all SNe except SN\,2017eaw (0.3 - 10 KeV) and SN\,2012aw (0.2 - 10 KeV). Where possible, the X-ray and radio luminosities are taken at the peak of the light curve and for others, the limits of peak flux density is used. We include only the Type IIP SNe with both X-ray and radio detection.}}
\end{figure*}

\subsection{Circumstellar Electron Temperature}
The circumstellar electron temperature is an uncertain parameter since the effect of SN radiation on the circumstellar gas needs to be estimated from the CSM models \citep{chevalier2003, chevalier2006b}. Calculations on Type IIL SNe suggest that the temperature at unit optical depth $T_{\rm cs}$ $\sim$ 3 $\times$ $10^{4}$ K for $\dot{M}_{-6}/w1$ = 3 and $T_{\rm cs}$ $\sim$ 1 $\times$ $10^{5}$ K for $\dot{M}_{-6}/w1$ = 10 \citep{lundqvist1988}. The mass-loss rate depends on $T_{e}$ as $\dot{M}$ $\propto$ $T_{e}^{0.68}$ \citep{weiler1986}, and hence the uncertainty in $T_{e}$ will cause significant errors in $\dot{M}$. Assuming an electron temperature of $T_{e}$ = 10$^{4}$ K, we derive the mass-loss rate of SN\,2004dj progenitor as $\dot{M} = (1.37 \pm 0.11) \times 10^{-6}\, M_{\odot}\rm yr^{-1}$ from our radio analysis and modelling. However, if we assume electron temperature $T_{e}$ = 10$^{5}$ K, the mass-loss rate will be $\dot{M} = (6.56 \pm 0.66) \times 10^{-6}\, M_{\odot}\rm yr^{-1}$. This is $\sim$ 20 times larger than the mass-loss rate derived from X-ray observations, i.e $(3.2 \pm 1.1) \times 10^{-7}\, M_{\odot}\rm yr^{-1}$ \citep{chakraborti2012}. It is also important to note that lower values of $\dot{M}$ favours IC cooling \citep{chevalier2006b,chakraborti2012}. If we use $\dot{M} = (6.56 \pm 0.66) \times 10^{-6}\, M_{\odot}\rm yr^{-1}$, the ratio of cooling timescales are $t_{\rm Comp}/t = 1.7 $  and $t_{\rm syn}/t = 1.9 $ on day 56 post explosion. Since both the ratios are much above 1, these numbers do not favour IC cooling. But there are several lines of evidences including the prominent IC X-ray component on day 56 post explosion \citep{chakraborti2012} showing that IC cooling is happening at the SN shock during this time as discussed in \S \ref{first-sub-section-cooling}. Thus the mass-loss rate derived for SN\,2004dj assuming a electron temperature of $T_{e}$ = $10^{5}$ K is too high to explain the observational signature of IC cooling. We suggest the electron temperature of the CSM to be $\sim$10$^{4}$ K rather than $10^{5}$ K for SN\,2004dj. We also note that \cite{misra2007} deduce the CSM temperature of another Type IIP SN\,2004et as 10$^{4}$ K from combined X-ray and radio data which has marginally higher wind density (see Table \ref{comparison-typeIIp}) and mass-loss rates \citep{misra2007} as that of SN\,2004dj.

\subsection{Comparison of SN\,2004dj with other Type IIP SNe}
Even though Type IIP is the most commonly observed variety of core-collapse SNe in the optical band, only few of them are known radio/X-ray emitters. SN\,2004dj being one of the best observed Type IIP SNe in the radio bands, here we compare the properties of SN\,2004dj with other radio/X-ray bright Type IIP SNe; SN\,1999em, SN\,1999gi, SN\,2002hh, SN\,2004et, SN\,2011ja, SN\,2013ej, SN\,2012aw and SN\,2017eaw. The physical parameters of these SNe along with the references are compiled in Table \ref{comparison-typeIIp}. We also plot the radio and X-ray luminosities of Type IIP SNe that was detected in both radio and X-ray bands in Figure 5.

The radio and X-ray luminosities of SN\,2004dj is comparable to that of SN\,1999em and SN\,2002hh (see Fig 5) and is suggestive of comparable CSM densities. The mass-loss rate of the progenitor of SN\,1999em and SN\,2002hh are $\sim$ 0.9 $\times$ 10$^{-6}$ $M_{\odot}\rm yr^{-1}$ and $\sim$ 1.2 $\times$ 10$^{-6}$ $M_{\odot}\rm yr^{-1}$ respectively \citep{chevalier2006b} for a CSM temperature of $T_{e}$ = $10^{4}$ K.  This is similar to the mass-loss rate derived for SN\,2004dj from our radio analysis (see Table \ref{comparison-typeIIp}). While the X-ray luminosity of SN\,2004et is comparable to that of SN\,2004dj, the radio luminosity of SN\,2004et is larger than ($\sim$ 3.5 times) SN\,2004dj and could be indicative of slightly larger wind densities. The mass-loss rate of SN\,2004et is derived from X-ray observations to be $\dot{M}$ $\sim$ 2 $\times$ 10$^{-6}$ $M_{\odot}\rm yr^{-1}$ \citep{misra2007} and $\dot{M}$ $\sim$ (1.6 - 1.8) $\times$ 10$^{-6}$ $M_{\odot}\rm yr^{-1}$ \citep{chevalier2006b}. These numbers are marginally larger than the mass-loss rate of SN\,2004dj derived from our analysis.

SN\,2012aw has a larger X-ray luminosity ($\sim$ 6 times) and radio luminosity \citep[$\sim$ 3 times;][]{yadav2014} as compared to SN\,2004dj (see Fig 5). The mass-loss rate is $\dot{M}$ $\sim$ 3.2 $\times$ 10$^{-6}$ $M_{\odot}\rm yr^{-1}$ from X-ray analysis \citep{kochanek2012} which is larger than the mass-loss rate of SN\,2004dj.

It is found that $\dot{M}$ depends on the metallicity of the regions of SNe as $\dot{M} \propto Z^{0.5}$ \citep{schaller1992,heger2003}. The metallicity of the regions of occurence of SN\,1999em \citep{smartt2003}, SN\,2004et \citep{li2005} and SN\,2004dj \citep{wang2005} are (1-2), (0.3-1) and $\sim$ 0.4 $Z_{\odot}$ respectively. This will have a minor effect in the $\dot{M}$ comparison discussed above. 

To summarize, the estimate of mass-loss rate of the progenitor star depends on various physical parameters including CSM electron temperature and metallicity of the cite of SN explosion. A fair comparison is not possible unless these quantities are well constrained by either observations or modelling. However, the mass-loss rates of Type IIP SNe progenitors deduced from radio and X-ray observations and modelling span over a range of $\sim$ (0.1 - 10)$\times$ 10$^{-6}$ $M_{\odot}\rm yr^{-1}$.

\section{Summary}
\label{sec:summary}
We carried out detailed radio observations of Type IIP supernova SN\,2004dj at frequencies ranging from 0.24 - 43 GHz at ages from 1.12 days to $\sim$ 12 years post discovery. We model the radio observations with standard mini-shell model \citep{chevalier1982}. Both FFA and SSA models fit with the data reasonably well and it is difficult to conclude either of them as the dominant absorption process from the modelled parameters. However, from the optical line velocity measurements, we infer FFA as the dominant absorption process which is consistent with the prediction by \cite{chevalier2006b}. The radio observations and modelling are consistent with the interaction of the SN with an outer ejecta density profile $\rho$ $\sim$ $r^{-11.4}$ with a circumstellar density field created by a pre-SN steady stellar wind. We derive the shock deceleration parameter $m$ $\sim$ 0.9 ($R$ $\sim$ $t^{m}$) indicative of a mildly decelerating blast wave. Assuming FFA to be the dominant absorption process, we derive the mass-loss rate of the progenitor star as $\dot{M}$ = (1.37 $\pm$ 0.11) $\times$ 10$^{-6}$\, $M_{\odot}\rm yr^{-1}$. The mass-loss rate derived from our observations are consistent with the theoretical predictions for a progenitor star of mass $M \sim 15 M_{\odot}$. The mass-loss rate derived from our analysis is $\sim$ 3 times larger than the value derived by \cite{chevalier2006b} for SN\,2004dj from early radio data. However, our mass-loss estimate is consistent with the range of RSG mass-loss rates of type IIP SNe \citep{chevalier2006b}. \cite{chakraborti2012} estimated the mass-loss rate, $\dot{M}$ = 3.2 $\times$ 10$^{-7}$  $M_{\odot}\rm yr^{-1}$ from X-ray emission measure and is $\sim$ 4 times smaller than the value derived from our analysis. We also present the evolution of radio spectral indices with 1.06, 1.4, 4.86, 8.46 GHz flux density measurements. The spectral indices steepen to values of $-$1 around day 50 and continues till $\sim$ day 125, especially at higher frequencies (4.86/8.46), suggestive of electron cooling. During this period, the optical light curve is in the plateau phase. We estimate the cooling time scale for both IC and synchrotron cooling and interpret the steepening as a signature of IC cooling at the SN shock.

 SN\,2004dj is the only Type IIP SN with radio data covering two order of magnitudes in time and frequency and this allowed us to study this SN as a prototype of Type IIP SNe. We compare the properties of SN\,2004dj with other radio/X-ray bright Type IIP SNe and find that SN\,2004dj is a normal type IIP SNe with very similar CSM properties as that of SN\,1999em and SN\,2002hh.

\acknowledgments
We thank the anonymous referee for the constructive comments. We thank Kurt Weiler for providing us the early VLA data for this supernova. We acknowledge Roger Chevalier for his comments on the manuscript.
P.C. acknowledges support from the Department of Science and Technology via SwaranaJayanti Fellowship award (file no.DST/SJF/PSA-01/2014-15). A.R. acknowledges Raja Ramana Fellowship of DAE, Govt of India. We thank the staff of the GMRT that made these observations possible. GMRT is run by the National Centre for Radio Astrophysics of the Tata Institute of Fundamental Research.
The National Radio Astronomy Observatory is a facility of the National Science Foundation operated under cooperative agreement by Associated Universities, Inc.
\software{AIPS \citep{van1996}}
\vspace{5mm}
\facilities{Giant Metrewave Radio Telescope, Karl J. Jansky Very Large Array}



\clearpage


\begin{thebibliography}{}

\bibitem[Argo et al.(2004)]{argo2004} Argo, M.~K., Muxlow, T.~W.~B., Beswick, R.~J., Pedlar, A., \& Marcaide, J.~M.\ 2004, \iaucirc, 8399, 3

\bibitem[Argo et al.(2005)]{argo2005} Argo, M.~K., Beswick, R.~J., Muxlow, T.~W.~B., et al.\ 2005, \memsai, 76, 565 
 

\bibitem[Argo et al.(2017)]{argo2017} Argo, M., Torres, M.~P., Beswick, R., \& Wrigley, N.\ 2017, The Astronomer's Telegram, 10472,  


\bibitem[Beswick et al.(2005)]{beswick2005} Beswick, R.~J., Muxlow, T.~W.~B., Argo, M.~K., et al.\ 2005, \apjl, 623, L21 

\bibitem[Beswick et al.(2005)]{beswick2005iauc} Beswick, R.~J., Fenech, D., Thrall, H., et al.\ 2005, \iaucirc, 8572, 1 

\bibitem[Bose \& Kumar(2014)]{bose2014} Bose, S., \& Kumar, B.\ 2014, \apj, 782, 98 


\bibitem[Bose et al.(2015)]{bose2015} Bose, S., Sutaria, F., Kumar, B., et al.\ 2015, \apj, 806, 160 

\bibitem[Chakraborti et al.(2013)]{chakraborti2013} Chakraborti, S., Ray, A., Smith, R., et al.\ 2013, \apj, 774, 30 


\bibitem[Chakraborti et al.(2016)]{chakraborti2016} Chakraborti, S., Ray, A., Smith, R., et al.\ 2016, \apj, 817, 22 

\bibitem[Chakraborti et al.(2012)]{chakraborti2012} Chakraborti, S., Yadav, N., Ray, A., et al.\ 2012, \apj, 761, 100 


\bibitem[Chandra \& Ray(2004)]{chandra2004} Chandra, P., \& Ray, A.\ 2004, \iaucirc, 8397, 3 

\bibitem[Chevalier(1982)]{chevalier1982} Chevalier, R.~A.\ 1982, \apj, 259, 302

\bibitem[Chevalier(1984)]{chevalier1984} Chevalier, R.~A.\ 1984, \apjl, 285, L63 


\bibitem[Chevalier(1996)]{chevalier1996} Chevalier, R.~A.\ 1996, Radio Emission from the Stars and the Sun, 93, 125 


\bibitem[Chevalier(1998)]{chevalier1998} Chevalier, R.~A.\ 1998, \apj, 499, 810 

\bibitem[Chevalier \& Fransson(2003)]{chevalier2003} Chevalier, R.~A., \& Fransson, C.\ 2003, Supernovae and Gamma-Ray Bursters, 598, 171 


\bibitem[Chevalier \& Fransson(2006)]{chevalier2006a} Chevalier, R.~A., \& Fransson, C.\ 2006, \apj, 651, 381 

\bibitem[Chevalier et al.(2006)]{chevalier2006b} Chevalier, R.~A., Fransson, C., \& Nymark, T.~K.\ 2006, \apj, 641, 1029 


\bibitem[Chugai et al.(2005)]{chugai2005} Chugai, N.~N., Fabrika, S.~N., Sholukhova, O.~N., et al.\ 2005, Astronomy Letters, 31, 792 


\bibitem[Chugai et al.(2007)]{chugai2007} Chugai, N.~N., Chevalier, R.~A., \& Utrobin, V.~P.\ 2007, \apj, 662, 1136 

\bibitem[de Jager et al.(1988)]{dejager1988} de Jager, C., Nieuwenhuijzen, H., \& van der Hucht, K.~A.\ 1988, \aaps, 72, 259 

\bibitem[Falk \& Arnett(1977)]{falk1977} Falk, S.~W., \& Arnett, W.~D.\ 1977, \apjs, 33, 515 


\bibitem[Filippenko(1997)]{filippenko1997} Filippenko, A.~V.\ 1997, \araa, 35, 309 

\bibitem[Fraser et al.(2012)]{fraser2012} Fraser, M., Maund, J.~R., Smartt, S.~J., et al.\ 2012, \apjl, 759, L13 


\bibitem[Fransson \& Bj{\"o}rnsson(1998)]{fransson1998} Fransson, C., \& Bj{\"o}rnsson, C.-I.\ 1998, \apj, 509, 861 


\bibitem[Freedman et al.(2001)]{freedman2001} Freedman, W.~L., Madore, B.~F., Gibson, B.~K., et al.\ 2001, \apj, 553, 47 


\bibitem[Grasberg et al.(1971)]{grasberg1971} Grasberg, E.~K., Imshenik, V.~S., \& Nadyozhin, D.~K.\ 1971, \apss, 28, 3 


\bibitem[Grefensetette et al.(2017)]{gref2017} Grefensetette, B., Harrison, F., \& Brightman, M.\ 2017, The Astronomer's Telegram, 10427,  


\bibitem[Heger et al.(2003)]{heger2003} Heger, A., Fryer, C.~L., Woosley, S.~E., Langer, N., \& Hartmann, D.~H.\ 2003, \apj, 591, 288 

\bibitem[Immler \& Brown(2012)]{immler2012} Immler, S., \& Brown, P.~J.\ 2012, The Astronomer's Telegram, 3995,  



\bibitem[Jura \& Kleinmann(1990)]{jura1990} Jura, M., \& Kleinmann, S.~G.\ 1990, \apjs, 73, 769 

\bibitem[Kochanek et al.(2012)]{kochanek2012} Kochanek, C.~S., Khan, R., \& Dai, X.\ 2012, \apj, 759, 20 



\bibitem[Kotak et al.(2005)]{kotak2005} Kotak, R., Meikle, P., van Dyk, S.~D., H{\"o}flich, P.~A., \& Mattila, S.\ 2005, \apjl, 628, L123 


\bibitem[Korcakova et al.(2005)]{korcakova2005} Korcakova, D., Mikulasek, Z., Kawka, A., et al.\ 2005, Information Bulletin on Variable Stars, 5605, 1 

\bibitem[Leonard et al.(2002)]{leonard2002} Leonard, D.~C., Filippenko, A.~V., Li, W., et al.\ 2002, \aj, 124, 2490 



\bibitem[Li et al.(2005)]{li2005} Li, W., Van Dyk, S.~D., Filippenko, A.~V., \& Cuillandre, J.-C.\ 2005, \pasp, 117, 121 


\bibitem[Li et al.(2006)]{li2006} Li, W., Van Dyk, S.~D., Filippenko, A.~V., et al.\ 2006, \apj, 641, 1060 

\bibitem[Lundqvist \& Fransson(1988)]{lundqvist1988} Lundqvist, P., \& Fransson, C.\ 1988, \aap, 192, 221 



\bibitem[Ma{\'{\i}}z-Apell{\'a}niz et al.(2004)]{maiz2004a} Ma{\'{\i}}z-Apell{\'a}niz, J., Bond, H.~E., Siegel, M.~H., et al.\ 2004, \apjl, 615, L113 

\bibitem[Maiz-Apellaniz(2004)]{maiz2004b} Maiz-Apellaniz, J.\ 2004, \iaucirc, 8385, 2 


\bibitem[Matzner \& McKee(1999)]{matzner1999} Matzner, C.~D., \& McKee, C.~F.\ 1999, \apj, 510, 379 


\bibitem[Maund \& Smartt(2005)]{maund2005a} Maund, J.~R., \& Smartt, S.~J.\ 2005, \mnras, 360, 288 

\bibitem[Maund et al.(2005)]{maund2005b} Maund, J.~R., Smartt, S.~J., \& Danziger, I.~J.\ 2005, \mnras, 364, L33 

\bibitem[Misra et al.(2007)]{misra2007} Misra, K., Pooley, D., Chandra, P., et al.\ 2007, \mnras, 381, 280 

\bibitem[Mouhcine et al.(2005)]{mouhcine2005} Mouhcine, M., Ferguson, H.~C., Rich, R.~M., Brown, T.~M., \& Smith, T.~E.\ 2005, \apj, 633, 810 



\bibitem[Nadyozhin(2003)]{nadyozhin2003} Nadyozhin, D.~K.\ 2003, \mnras, 346, 97 

\bibitem[Nakano et al.(2004)]{nakano2004} Nakano, S., Itagaki, K., Bouma, R.~J., Lehky, M., \& Hornoch, K.\ 2004, \iaucirc, 8377, 1 

\bibitem[Nayana \& Chandra(2017)]{nayana2017} Nayana, A.~J., \& Chandra, P.\ 2017, The Astronomer's Telegram, 10534,  


\bibitem[Nieuwenhuijzen \& de Jager(1990)]{nieuwenhuijzen1990} Nieuwenhuijzen, H., \& de Jager, C.\ 1990, \aap, 231, 134 


\bibitem[Patat et al.(2004)]{patat2004} Patat, F., Benetti, S., Pastorello, A., Filippenko, A.~V., \& Aceituno, J.\ 2004, \iaucirc, 8378, 1 

\bibitem[Pooley \& Lewin(2002)]{pooley2002iauc} Pooley, D., \& Lewin, W.~H.~G.\ 2002, \iaucirc, 8024, 2 

\bibitem[Pooley et al.(2002)]{pooley2002} Pooley, D., Lewin, W.~H.~G., Fox, D.~W., et al.\ 2002, \apj, 572, 932 


\bibitem[Pooley \& Lewin(2004)]{pooley2004} Pooley, D., \& Lewin, W.~H.~G.\ 2004, \iaucirc, 8390, 1 

\bibitem[Pilyugin et al.(2004)]{pilyugin2004} Pilyugin, L.~S., V{\'{\i}}lchez, J.~M., \& Contini, T.\ 2004, \aap, 425, 849 

\bibitem[Reimers(1977)]{reimers1977} Reimers, D.\ 1977, \aap, 61, 217 


\bibitem[Schaller et al.(1992)]{schaller1992} Schaller, G., Schaerer, D., Meynet, G., \& Maeder, A.\ 1992, \aaps, 96, 269 

\bibitem[Schlegel(2001)]{schlegel2001} Schlegel, E.~M.\ 2001, \apjl, 556, L25 



\bibitem[Smartt et al.(2003)]{smartt2003} Smartt, S.~J., Maund, J.~R., Gilmore, G.~F., et al.\ 2003, \mnras, 343, 735 

\bibitem[Smartt et al.(2004)]{smartt2004} Smartt, S.~J., Maund, J.~R., Hendry, M.~A., et al.\ 2004, Science, 303, 499 


\bibitem[Smartt et al.(2009)]{smartt2009a} Smartt, S.~J., Eldridge, J.~J., Crockett, R.~M., \& Maund, J.~R.\ 2009, \mnras, 395, 1409 

\bibitem[Smartt(2009)]{smartt2009b} Smartt, S.~J.\ 2009, \araa, 47, 63

\bibitem[Smartt et al.(2004)]{smartt2004} Smartt, S.~J., Maund, J.~R., Hendry, M.~A., et al.\ 2004, Science, 303, 499 


\bibitem[Smith et al.(2011)]{smith2011} Smith, N., Li, W., Filippenko, A.~V., \& Chornock, R.\ 2011, \mnras, 412, 1522 



\bibitem[Stockdale et al.(2004)]{stockdale2004} Stockdale, C.~J., Sramek, R.~A., Weiler, K.~W., et al.\ 2004, \iaucirc, 8379, 1 


\bibitem[Sugerman et al.(2005)]{sugerman2005} Sugerman, B., Seeds Collaboration, Sings Legacy, P., \& van Dyk, S.\ 2005, \iaucirc, 8489, 2 

\bibitem[Van Dyk et al.(1994)]{vandyk1994} Van Dyk, S.~D., Weiler, K. W., Sramek, R. A., Rupen, M. P., \& Panagia, N. \ 1994, \ ApJ, 432, L115


\bibitem[Van Dyk et al.(2003)]{vandyk2003} Van Dyk, S.~D., Li, W., \& Filippenko, A.~V.\ 2003, \pasp, 115, 1289 

\bibitem[Van Dyk et al.(2012)]{vandyk2012} Van Dyk, S.~D., Cenko, S.~B., Poznanski, D., et al.\ 2012, \apj, 756, 131

\bibitem[van Moorsel et al.(1996)]{van1996} van Moorsel, G., Kemball, A., \& Greisen, E.\ 1996, Astronomical Data Analysis Software and Systems V, 101, 37 


\bibitem[van Loon et al.(2005)]{van2005} van Loon, J.~T., Cioni, M.-R.~L., Zijlstra, A.~A., \& Loup, C.\ 2005, \aap, 438, 273 


\bibitem[Vink{\'o} et al.(2006)]{vinko2006} Vink{\'o}, J., Tak{\'a}ts, K., S{\'a}rneczky, K., et al.\ 2006, \mnras, 369, 1780 

\bibitem[Wang et al.(2005)]{wang2005} Wang, X., Yang, Y., Zhang, T., et al.\ 2005, \apjl, 626, L89 

\bibitem[Weiler et al.(1986)]{weiler1986} Weiler, K.~W., Sramek, R.~A., Panagia, N., van der Hulst, J.~M., \& Salvati, M.\ 1986, \apj, 301, 790 

\bibitem[Weiler et al.(2002)]{weiler2002} Weiler, K.~W., Panagia, N., Montes, M.~J., \& Sramek, R.~A.\ 2002, \araa, 40, 387 

\bibitem[Yadav et al.(2014)]{yadav2014} Yadav, N., Ray, A., Chakraborti, S., et al.\ 2014, \apj, 782, 30 



\bibitem[Zhang et al.(2006)]{zhang2006} Zhang, T., Wang, X., Li, W., et al.\ 2006, \aj, 131, 2245 




\end{thebibliography}
\end{document}